\documentclass{elsart}

\usepackage{graphicx}

\usepackage{epsfig}
\usepackage{amssymb}

\newcommand{\be}{\begin{equation}}
\newcommand{\ee}{\end{equation}}
\newcommand{\ba}{\begin{eqnarray}}
\newcommand{\ea}{\end{eqnarray}}

\begin{document}

\begin{frontmatter}

\title{Gapless color superconductivity at zero and at finite temperature}

\author{Mei Huang\thanksref{Tsinghua-thank}}
\ead{huang@th.physik.uni-frankfurt.de}
\thanks[Tsinghua-thank]{On leave from Physics Department, 
Tsinghua University, Beijing 100084, China.}
and
\author{Igor Shovkovy\thanksref{BITP-thank}\corauthref{cor-author}}
\ead{shovkovy@th.physik.uni-frankfurt.de}
\thanks[BITP-thank]{On leave from Bogolyubov Institute for
Theoretical Physics, 03143, Kiev, Ukraine.}

\address{Institut f\"{u}r Theoretische Physik, 
        J.W. Goethe-Universit\"{a}t,
        D-60054 Frankurt/Main, Germany}

\corauth[cor-author]{Corresponding author.
Full postal address:\\
\hspace*{6.5cm}Institut f\"{u}r Theoretische Physik,\\
\hspace*{6.5cm}J.W. Goethe-Universit\"{a}t, \\
\hspace*{6.5cm}Postfach 11 19 32, \\
\hspace*{6.5cm}D-60054 Frankurt/Main, Germany\\
\hspace*{2cm}Tel.:(+49) (69) 798-23310, Fax: (+49) (69) 798-28350}

\date{\today}

\begin{abstract}

We present a detailed study of the recently proposed gapless color
superconductivity in neutral two-flavor quark matter in 
$\beta$-equilibrium at zero as well as at finite temperature. We 
clarify the nature of the fermionic quasiparticles, and discuss the 
physical properties of this gapless phase of quark matter. The 
appearance of two ``secondary" spin-1 condensates in the gapless 
ground state is pointed out. The gluon screening properties are
also discussed. It is found that finite temperature properties of 
the gapless color superconductor are rather unusual. One of the most 
striking results is that the ratio of the critical temperature to the gap 
at zero temperature is not a universal number. Moreover, this ratio
could become arbitrarily large at some values of the coupling
constant. In addition, the value of the gap parameter is not a 
monotonic function of temperature, although the phase transition 
is a second order phase transition. In an extreme case, in fact,
a nonzero finite temperature gap can appear even if its zero 
temperature value is vanishing.

\end{abstract}

\end{frontmatter}

{\bf e-Print archive number: {\tt hep-ph/0307273}}

\newpage

\section{Introduction}
\label{intor}

Sufficiently cold and dense quark matter is a color superconductor
\cite{old}. The corresponding ground state is characterized by a
condensate of Cooper pairs made of quarks. Since the latter carry color
charge, the SU(3)$_{c}$ color gauge group of strong interactions is
partially or completely broken through the Anderson-Higgs mechanism. At
asymptotic densities, this phenomenon was studied from first principles in
Refs.~\cite{weak,PR-sp1,weak-cfl}. These studies of dense QCD, as well as
studies of various phenomenologically motivated models
\cite{cs,cfl,crystal,spin-1,spin-1-Meissner}, suggest that dense quark 
matter at high baryon density has a very rich phase structure, consisting 
of many different phases. 

It is natural to expect that some color superconducting phases may exist
in the interior of compact stars. The estimated central densities of such
stars might be as large as $10 \rho_{0}$ (where $\rho_{0}\approx 0.15$   
fm$^{-3}$ is the saturation density), while their temperatures are in the 
range of tens of keV. This could provide ideal conditions for the diquark 
Cooper pairing of color superconductivity.

Matter in the bulk of compact stars should be neutral with respect to
electrical as well as color charges. Also, such matter should remain in
$\beta$-equilibrium. Satisfying these requirements may impose nontrivial
relations between the chemical potentials of different quarks. In turn,
such relations could substantially influence the pairing dynamics between
quarks, for instance, by suppressing some color superconducting phases
and by favoring others. For example, it was argued in 
Ref.~\cite{absence2sc}, that the mixture of the 2-flavor color 
superconducting (2SC) phase and normal strange quarks is less favorable 
than the color-flavor-locked (CFL) phase if the charge neutrality condition 
is enforced (note that the strange quark mass was chosen too small to 
allow the appearance of a pure 2SC phase). A similar conclusion was also 
reached in Ref.~\cite{neutral_steiner}. 

Recently, it was proposed that neutral two-flavor quark matter in
$\beta$-equilibrium can have a rather unusual, stable ground state 
called a gapless color superconductor (g2SC) \cite{SH}. A similar 
phase of gapless color superconductivity could also appear in strange 
quark matter when the baryon number density is kept fixed \cite{GLW} 
[for earlier studies see Ref.~\cite{gaplessCFL} where an unstable 
gapless phase was considered]. This gapless strange quark matter may, 
in principle, be realized inside the so-called strangelets. It was 
also suggested that closely related phases may exist in two component 
mixtures of cold fermionic atoms \cite{WilLiu}. It is fair to mention
that, in both quark and cold atom systems, the ground state could 
also be given by some mixed phases \cite{neutral_buballa,SHH,Bed}
provided the surface tension effects are not too strong. 

While the symmetry of the g2SC ground state is the same as that of 
the conventional 2-flavor color superconductor, the spectrum of 
the fermionic quasiparticles is different. In particular, two out of 
four gaped quasiparticles of the conventional 2SC phase become gapless 
in the g2SC phase. In addition, the number densities of the pairing quarks 
in g2SC at zero temperature are not equal \cite{SH}. For example, the 
density of red (green) up quarks is different from the density of green 
(red) down quarks. This is in contrast to the conventional ``enforced" 
pairing \cite{enforced} where the corresponding densities are equal.

It is possible that the g2SC quark matter exists inside central regions 
of compact stars. This would be quite natural since this phase of matter 
is neutral with respect to electrical and color charges, and satisfies the
$\beta$-equilibrium condition by construction. Also, this phase is more 
stable than normal quark matter under similar conditions \cite{SH}. While 
suggestive, these arguments alone are not sufficient to make a conclusive 
statement regarding the chances of the g2SC phase to appear inside stars. 
In order to study this issue properly, one should consider the competition 
of the g2SC phase with all other possible phases of quark/hadronic matter. 
However, the limited knowledge of the strongly coupled dynamics in QCD 
at realistic densities does not allow such a systematic quantitative 
study.

In this paper, we study properties of the g2SC phase at finite temperature. 
Our interest in such properties is driven by the expectation that the g2SC 
phase of quark matter may already appear at a relatively early (protoneutron) 
stage of the compact star evolution when the temperature of the stellar 
core is quite high. If this is the case, the knowledge of 
finite temperature properties of quark matter would be needed to 
understand the stellar properties and, possibly, even to predict 
observational signatures of the new state of matter inside stars. 

This paper is organized as follows. In the next section, we set our
notation by presenting a model of quark matter, and spell out the main
assumptions in the analysis. Then, in Sec.~\ref{T=0} we discuss in detail
the solution to the gap equation and the condition of the charge
neutrality in the two-flavor quark model at zero temperature. In addition
to reproducing the numerical results of our original short Letter
\cite{SH}, we also present many analytical results that have not been
included in Ref.~\cite{SH}. At zero temperature, the screening properties 
of gluons in the g2SC phase, as well as the possible appearance of two 
additional spin-1 condensates, are deduced from the qualitative nature 
of the low energy spectrum of quasiparticles. The finite temperature 
properties of the g2SC phase are studied in Sec.~\ref{finite-T}. Among 
main results, we find that (i) the phase transition between the g2SC 
phase and normal quark matter is a second order phase transition, 
(ii) the ratio of the critical temperature and the zero temperature 
gap is not a universal number in the gapless phase, and (iii) a nonzero 
finite temperature gap may appear even if its zero temperature value 
is vanishing. Sec.~\ref{conclusion} contains our conclusions and a 
brief discussion of the properties of the g2SC phase as well as their 
applications.

\section{Model}
\label{model}

\subsection{Lagrangian density and parameters of the model}

Let us start our analysis by discussing the quark model that we use in the
following. We assume that the strange quark is sufficiently heavy and does
not appear at intermediate baryon densities under consideration (e.g.,
this might correspond to the quark chemical potential $\mu \equiv
\mu_{B}/3$ in the range between about $350$ and $450$ MeV). Therefore, we
use the simplest SU(2) Nambu--Jona-Lasinio (NJL) model that was proposed
in Ref.~\cite{huang_2sc}. The explicit form of the Lagrangian density
reads: 
\begin{eqnarray}
\label{lagr}
L & = &\bar{q}(i\gamma^{\mu}\partial_{\mu}-m_0)q + 
 G_S\left[(\bar{q}q)^2 + (\bar{q}i\gamma_5{\bf \vec{\tau}}q)^2\right] 
\nonumber \\
 &&+G_D\left[(i \bar{q}^C  \varepsilon  \epsilon^{b} \gamma_5 q )
 (i \bar{q} \varepsilon \epsilon^{b} \gamma_5 q^C)\right],
\end{eqnarray}
where $q^C=C {\bar q}^T$ is the charge-conjugate spinor and $C=i\gamma^2
\gamma^0$ is the charge conjugation matrix. The quark field $q \equiv
q_{i\alpha}$ is a four-component Dirac spinor that carries flavor
($i=1,2$) and color ($\alpha=1,2,3$) indices. The Pauli matrices
are denoted by ${\vec \tau} =(\tau^1,\tau^2, \tau^3)$, while
$(\varepsilon)^{ik} \equiv \varepsilon^{ik}$ and $(\epsilon^b)^{\alpha
\beta} \equiv \epsilon^{\alpha \beta b}$ are the antisymmetric tensors 
in the flavor and color spaces, respectively. We also introduce a 
momentum cutoff $\Lambda$, and two independent coupling constants in 
the scalar quark-antiquark and scalar diquark channels, $G_S$ and $G_D$.

The values of the parameters in the NJL model are chosen by fitting
the pion decay constant $f_{\pi}=93$ MeV and the chiral condensate
$\langle {\bar u} u \rangle^{1/3}  = \langle {\bar d} d \rangle^{1/3} 
=-250$ MeV. In the case of $m_0\neq 0$, the additional parameter could 
also be fixed by fitting the value of the pion mass. In the rest of this 
paper, however, we consider only the chiral limit with $m_{0}=0$. Then
the two model parameters, i.e., the coupling constant and the cutoff, 
are chosen as follows: $G_S=5.0163$ GeV$^{-2}$ and $\Lambda=653.3$ MeV 
\cite{huang_2sc,SKP}. Without loss of generality, we set the strength 
of the diquark coupling $G_D$ to be proportional to the quark-antiquark 
coupling constant, i.e., $G_D = \eta G_S$ with a typical number for 
$\eta$ being around $0.75$ \cite{SH}.

When considering the chiral limit in the following, we shall find that 
the constituent quark mass is zero in the model at hand in the g2SC 
phase. In reality, however, this may not be the case. At the same time, 
since the appearance of the g2SC phase is a Fermi surface phenomenon, 
our assumption should not be a strong limitation. Of course, nonzero 
quark masses should affect the Fermi momenta of different quark species.
In turn, this could change slightly the mismatch parameter $\delta\mu$
in our model (see below), and introduce some minor changes in the analysis 
\cite{misra}. However, to large extent, the results would not change.

\subsection{Thermodynamic potential in $\beta$-equilibrium}

In $\beta$-equilibrium, the diagonal matrix of quark chemical potentials
is given in terms of quark, electrical and color chemical potentials,
\begin{equation} 
\mu_{ij, \alpha\beta}= (\mu \delta_{ij}- \mu_e Q_{ij})
\delta_{\alpha\beta} + \frac{2}{\sqrt{3}}\mu_{8} \delta_{ij} 
(T_{8})_{\alpha \beta},
\end{equation} 
where $Q$ and $T_8$ are generators of U(1)$_{em}$ of electromagnetism
and the U(1)$_{8}$ subgroup of the color gauge group. The explicit
expressions for the quark chemical potentials read
\begin{eqnarray}
\mu_{ur} =\mu_{ug} =\mu -\frac{2}{3}\mu_{e} +\frac{1}{3}\mu_{8}, \\
\mu_{dr} =\mu_{dg} =\mu +\frac{1}{3}\mu_{e} +\frac{1}{3}\mu_{8}, \\
\mu_{ub} =\mu -\frac{2}{3}\mu_{e} -\frac{2}{3}\mu_{8}, \\
\mu_{db} =\mu +\frac{1}{3}\mu_{e} -\frac{2}{3}\mu_{8}. 
\end{eqnarray}
One should notice that, in general, there are two mutually commuting 
color charges related to the generators $T_3$ and $T_8$ of the 
SU(3)$_{c}$ group. Therefore, one could introduce two chemical potentials 
for these two different color charges. However, we require that quark 
matter in the 2SC ground state remains invariant under the SU(2)$_{c}$ 
color gauge subgroup. This condition makes the introduction of the 
second nontrivial color chemical potential $\mu_{3}$ unnecessary.

In the mean field approximation, the finite temperature effective 
potential for quark matter in $\beta$-equilibrium with electrons takes 
the form:
\begin{eqnarray} 
\label{pot}
\Omega &=& \Omega_{0}
-\frac{1}{12\pi^2}\left(\mu_{e}^{4}+2\pi^{2}T^{2}\mu_{e}^{2}
+\frac{7\pi^{4}}{15} T^{4} \right) +\frac{(m-m_0)^2}{4G_S} \nonumber\\
&+&\frac{\Delta^2}{4G_D}
-\sum_{a} \int\frac{d^3 p}{(2\pi)^3} \left[E_{a}
+2 T\ln\left(1+e^{-E_{a}/T}\right)\right],
\end{eqnarray}
where $\mu_{e}$ is the electron chemical potential and $\Omega_{0}$ is a
constant added to make the pressure of the vacuum zero. For simplicity, we
substituted the zero value of the electron mass which is sufficient
for the purposes of the current study. The sum in the second line of
Eq.~(\ref{pot}) runs over all (6 quark and 6 antiquark)
quasiparticles. The explicit dispersion relations and the degeneracy 
factors of the quasiparticles read
\begin{eqnarray}
E_{ub}^{\pm} &=& E(p) \pm \mu_{ub} , \hspace{26.6mm} [\times 1]
\label{disp-ub} \\
E_{db}^{\pm} &=& E(p) \pm \mu_{db} , \hspace{26.8mm} [\times 1]
\label{disp-db}\\
E_{\Delta^{\pm}}^{\pm} &=& E_{\Delta}^{\pm}(p) \pm  \delta \mu .
\hspace{25.5mm} [\times 2]
\label{2-degenerate}
\end{eqnarray}
Here we introduced the following shorthand notation:
\begin{eqnarray}
E(p) &\equiv& \sqrt{{\bf p}^2+m^2}, \\
E_{\Delta}^{\pm}(p) &\equiv& 
\sqrt{[E(p) \pm \bar{\mu}]^2 +\Delta^2},\\
\bar{\mu} &\equiv& 
\frac{\mu_{ur} +\mu_{dg}}{2} 
=\frac{\mu_{ug}+\mu_{dr}}{2}
=\mu-\frac{\mu_{e}}{6}+\frac{\mu_{8}}{3}, \label{mu-bar}\\
\delta\mu &\equiv& 
 \frac{\mu_{dg}-\mu_{ur}}{2}
=\frac{\mu_{dr}-\mu_{ug}}{2}
=\frac{\mu_{e}}{2}. \label{delta-mu}
\end{eqnarray}
The knowledge of the thermodynamic potential is sufficient to calculate 
all other thermodynamic properties of quark matter in equilibrium. 
In the model at hand, the physical thermodynamic potential that 
determines the pressure of quark matter, $\Omega_{\rm phys} =-P$,
is obtained from $\Omega$ in Eq.~(\ref{pot}) after substituting 
$\mu_{8}$, $\mu_{e}$, $m$ and $\Delta$ that solve the color and 
electrical charge neutrality conditions, i.e.,
\begin{equation}
n_{8}\equiv
-\frac{\partial \Omega}{\partial \mu_{8}}=0,  \quad \mbox{and} \quad 
n_{Q}\equiv
-\frac{\partial \Omega}{\partial \mu_{e}}=0,
\label{Q=0}
\end{equation}
as well as the gap equations,
\begin{equation}
\frac{\partial \Omega}{\partial m}=0, \quad \mbox{and} \quad
\frac{\partial \Omega}{\partial \Delta}=0,
\label{gap-eqs}
\end{equation}
respectively.
In the following two subsections, we shall derive the explicit expressions for 
the charge neutrality conditions and the gap equations.  

\subsection{Electrical and color charge densities}

By enforcing only the charge neutrality conditions in Eq.~(\ref{Q=0}),
one can construct the effective potential of the neutral quark matter 
as a function of two order parameters $m$ and $\Delta$. The subsequent
minimization of such a potential determines the physical values of the 
Dirac mass of quarks and the value of the superconducting gap in the 
ground state of neutral quark matter.

From the potential in Eq.~(\ref{pot}), we derive the expressions for 
the electrical charge density $n_Q$ and the color charge density
$n_8$ by taking partial derivatives with respect to $\mu_e$ and 
$\mu_8$, respectively. As a result, we arrive at the following 
explicit expressions:
\begin{eqnarray} 
n_8 &=& \frac{4}{3}\int\frac{d^3{\bf p}}{(2\pi)^3}
\Bigg[ - \frac{E-\bar\mu}{E_{\Delta}^-}
[1- {\tilde f}(E_{\Delta^+}^-) 
- {\tilde f}(E_{\Delta^-}^- ) ] \nonumber \\
&  & +  \frac{E+\bar\mu}{E_{\Delta}^+}
[1- {\tilde f}(E_{\Delta^+}^+)
- {\tilde f}(E_{\Delta^-}^+ ) ] \nonumber \\
& & + {\tilde f}(E_{ub}^+) - {\tilde f}(E_{ub}^-)
+ {\tilde f}(E_{db}^+) - {\tilde f}(E_{db}^-) \Bigg],
\label{n8}
\end{eqnarray}
and
\begin{eqnarray}
n_Q &= & 
-\frac{1}{2} n_8 +\frac{\mu_e^3}{3\pi^2} + \frac{1}{3}\mu_e T^2
+2 \int\frac{d^3{\bf p}}{(2\pi)^3} \Bigg[ 
{\tilde f}(E_{ub}^+) - {\tilde f}(E_{ub}^-) \nonumber \\
& &
-{\tilde f}(E_{\Delta^+}^-) - {\tilde f}(E_{\Delta^+}^+)
 + {\tilde f}(E_{\Delta^-}^+) + {\tilde f} (E_{\Delta^-}^-)
\Bigg] .
\label{nQ}
\end{eqnarray}
Here the standard Fermi distribution function has been introduced,
$ {\tilde f}(E)=1/[1+\exp(E/T)]$. From Eq.~(\ref{nQ}) we notice 
that the linear combination of charge densities $n_{Q}+\half n_{8}$ is  
given by a rather simple expression. In this particular combination,
the electrical and color charges of blue down quarks cancel. Had the 
dispersion relations in Eq.~(\ref{2-degenerate}) been dependent on 
$\bar\mu$ only, the charge cancellation from the corresponding 
quasiparticles would also be exact. It is the presence of the 
mismatch parameter $\delta\mu$ that leads to a nonvanishing 
contribution. As we shall see, the latter survives even at zero 
temperature in the g2SC phase ($\delta\mu>\Delta$), while disappear 
in the 2SC phase ($\delta\mu<\Delta$).

In the following, we shall require both charge densities in 
Eqs.~(\ref{n8}) and (\ref{nQ}) to be zero. Such a requirement is 
necessary in order to have a realistic construction of dense 
quark matter that could appear in the bulk of a compact star. 
Indeed, without enforcing the neutrality, the repulsive Coulomb 
force would clearly overwhelm the gravity preventing the formation 
of a stable star.

Before concluding this subsection, we mention that it also 
makes sense to study quark matter at fixed values of chemical 
potentials without imposing the conditions of local neutrality. 
Such studies are particularly appropriate when globally neutral 
mixed phases are considered \cite{neutral_buballa,SHH}.  
In this paper, however, we do not discuss mixed phases, but 
concentrate our attention exclusively on locally neutral (homogeneous) 
phases of quark matter in $\beta$-equilibrium.

\subsection{Gap equations for order parameters}

Starting from the potential in Eq.~(\ref{pot}), we derive an
explicit form of the gap equations. In particular, the equation 
for the Dirac mass reads
\begin{eqnarray}
 m - m_0 &=& 4G_S 
\int\frac{d^3{\bf p}}{(2\pi)^3}\frac{m}{E} 
\Bigg[ 2\frac{E-\bar\mu}{E_{\Delta}^-}[1- {\tilde f}(E_{\Delta^+}^-) 
- {\tilde f}(E_{\Delta^-}^- ) ] \nonumber \\
&  &  + 2 \frac{E+\bar\mu}{E_{\Delta}^+}
[1- {\tilde f}(E_{\Delta^+}^+)
- {\tilde f}(E_{\Delta^-}^+ ) ]  \nonumber \\
&  & + 
2- {\tilde f}(E_{ub}^+) - {\tilde f}(E_{ub}^-) 
- {\tilde f}(E_{db}^+) - {\tilde f}(E_{db}^-)
\Bigg],
\end{eqnarray}
In the chiral limit ($m_0=0$), the trivial solution $m=0$ corresponds 
to a chirally symmetric phase of quark matter, while a nontrivial 
solution $m \neq 0$ corresponds to a phase with spontaneously 
broken chiral symmetry.

Similarly, we derive the gap equation for the diquark condensate,
\begin{eqnarray}
 \Delta &=& 8G_D\int\frac{d^3{\bf p}}{(2\pi)^3} 
\Bigg[ \frac{\Delta}{E_{\Delta}^-}
[1- {\tilde f}(E_{\Delta^+}^-) - {\tilde f}(E_{\Delta^-}^- )]
\nonumber \\  
&&+ \frac{\Delta}{E_{\Delta}^+}
[1- {\tilde f}(E_{\Delta^+}^+)- {\tilde f}(E_{\Delta^-}^+ )] 
\Bigg].
\label{gap-T}
\end{eqnarray}
This equation can also have trivial as well as nontrivial 
solutions for $\Delta$. In the latter case, the corresponding
ground state is a color superconductor.

\subsection{Number densities of quarks}

According to one of the criteria of the g2SC phase, the densities 
of the quark species that participate in pairing dynamics are not 
equal at zero temperature \cite{SH}. This is in contrast to regular 
pairing in the conventional gaped color superconductors \cite{enforced}. 

Even before studying the effect of the Cooper pairing on the quark 
densities, we should discuss how to derive the formal expressions 
for these densities. One appreciates immediately, of course, that the 
definitions of such densities are unambiguous only if the chemical 
potentials of all six types (2 flavors times 3 colors) of quarks 
are treated as independent parameters. This is exactly how we proceed. 
The quark densities are obtained by taking the partial 
derivatives of the potential $\Omega$ in Eq.~(\ref{pot}) with 
respect to these independent parameters. At the end, the relations 
between the chemical potentials required by the $\beta$-equilibrium are 
imposed. The only point that needs some additional explanation is 
how to interpret the dependence of the thermodynamic potential 
in Eq.~(\ref{pot}) on the parameters $\bar\mu$ and $\delta\mu$, 
entering through the quasiparticle energies in 
Eq.~(\ref{2-degenerate}). It turns out that $\bar\mu$ and 
$\delta\mu$ should be replaced by $\half(\mu_{dg}+\mu_{ur})$ and 
$\half(\mu_{dg}-\mu_{ur})$, respectively, in one set of the double 
degenerate modes, and by $\half(\mu_{dr}+\mu_{ug})$ and 
$\half(\mu_{dr}-\mu_{ug})$ in the other set, see Eqs.~(\ref{mu-bar}) 
and (\ref{delta-mu}).

Because of the residual $SU(2)_c$ symmetry in the ground state of 
quark matter, the densities of the same flavor, red and green quarks 
are equal in the g2SC ground state. For example, the densities of 
the up quarks participating in the Cooper pairing read
\begin{eqnarray}
 n_{ur} = n_{ug} & = & \int\frac{d^3{\bf p}}{(2\pi)^3} 
\Bigg[\frac{E+\bar\mu}{E_{\Delta}^+}
[1- {\tilde f}(E_{\Delta^+}^+)
    -{\tilde f}(E_{\Delta^-}^+ ) ] \nonumber \\
 &  &- \frac{E-\bar\mu}{E_{\Delta}^-}
[1- {\tilde f}(E_{\Delta^+}^-)
- {\tilde f}(E_{\Delta^-}^- ) ]\nonumber \\
&  & + {\tilde f}(E_{\Delta^+}^-) + {\tilde f}(E_{\Delta^+}^+)
 - {\tilde f}(E_{\Delta^-}^+) - {\tilde f}(E_{\Delta^-}^-)\Bigg].
\end{eqnarray}
Similarly, for the densities of the down quarks, we get
\begin{eqnarray}
 n_{dg} = n_{dr} & = & \int\frac{d^3{\bf p}}{(2\pi)^3} 
   \Bigg[\frac{E+\bar\mu}{E_{\Delta}^+}
[1- {\tilde f}(E_{\Delta^+}^+)
    -{\tilde f}(E_{\Delta^-}^+ )] \nonumber \\
&&-\frac{E-\bar\mu}{E_{\Delta}^-}[1- {\tilde f}(E_{\Delta^+}^-)
- {\tilde f}(E_{\Delta^-}^- ) ]\nonumber \\
&  & - {\tilde f}(E_{\Delta^+}^-) - {\tilde f}(E_{\Delta^+}^+)
 + {\tilde f}(E_{\Delta^-}^+) + {\tilde f}(E_{\Delta^-}^-)\Bigg]. 
\end{eqnarray}
As we see, the densities of the up and down quarks participating in 
the Cooper pairing are not equal. In fact, the difference of the 
densities is given by
\begin{eqnarray}
n_{dg} - n_{ur} &=& n_{dr} - n_{ug} \nonumber \\
 & = & 2 \int\frac{d^3{\bf p}}{(2\pi)^3}
\left[{\tilde f}(E_{\Delta^-}^-)-{\tilde f}(E_{\Delta^+}^-) 
+{\tilde f}(E_{\Delta^-}^+)-{\tilde f}(E_{\Delta^+}^+)\right],
\label{n_dg-n_ur}
\end{eqnarray}
which is always nonzero at finite temperature provided the mismatch 
parameter $\delta\mu$ is nonzero. It is even more important for us here 
that this difference is nonzero at zero temperature in the g2SC phase 
of quark matter,
\begin{eqnarray}
\left.\left(
n_{dg} - n_{ur} \right)\right|_{T=0} &=& \left.\left(
n_{dr} - n_{ug} \right)\right|_{T=0} \nonumber \\
& = & 
\theta\left(\delta\mu-\Delta\right)
\frac{2}{3\pi^2} \sqrt{(\delta\mu)^2-\Delta^2}
\left(3\bar\mu^2+(\delta\mu)^2-\Delta^2\right).
\label{n_dg-n_ur-T0}
\end{eqnarray}
This is in contrast to the 2SC phase ($\delta\mu <\Delta$) where
this difference is zero in agreement with the arguments of 
Ref.~\cite{enforced}. 

The unpaired blue quarks, $u_b$ and $d_b$, are singlet states with 
respect to SU(2)$_c$ symmetry of the ground state. The densities of
these quarks are 
\begin{eqnarray}
n_{ub}= 2 \int\frac{d^3{\bf p}}{(2\pi)^3}
\left[ {\tilde f}(E_{ub}^-)-{\tilde f}(E_{ub}^+) \right]
\simeq \frac{\mu_{ub}}{3}\left(\frac{\mu_{ub}^2}{\pi^2} + T^2\right),
\end{eqnarray}
for the up quarks, and
\begin{eqnarray}
n_{db}= 2 \int\frac{d^3{\bf p}}{(2\pi)^3}
\left[ {\tilde f}(E_{db}^-)-{\tilde f}(E_{db}^+) \right]
\simeq \frac{\mu_{db}}{3}\left(\frac{\mu_{db}^2}{\pi^2} + T^2\right),
\end{eqnarray}
for the down quarks, respectively.

\section{Gapless 2SC at zero temperature}
\label{T=0}

Here we present a detailed study of the zero temperature properties 
of dense two-flavor quark matter. A brief outline of this study was 
presented in Ref.~\cite{SH}. In this paper, we include additionally 
many new results (e.g., regarding the gluon properties in g2SC phase 
of quark matter, and the appearance of spin-1 color superconducting 
gaps on top of the g2SC phase) that were outside the scope of the 
original publication \cite{SH}. 

At zero temperature, the effective potential (\ref{pot}) for quark 
matter in $\beta$-equilibrium with electrons takes the following
simple form \cite{SH}:
\begin{equation}
\label{potential}
\Omega = \Omega_{0}-\frac{\mu_e^4}{12 \pi^2}
+\frac{m^2}{4G_S}+\frac{\Delta^2}{4G_D}
- \sum_{a} \int\frac{d^3 p}{(2\pi)^3} |E_{a}|.
\end{equation}
Here, without loss of generality, we assume that $m_0=0$ (chiral 
limit). In this case, the constituent quark mass is also zero in
the color superconducting phase of matter at sufficiently large
baryon chemical potential. Thus, we substitute $m=0$ below. After 
making use of the quasiparticle energies in 
Eqs.~(\ref{disp-ub})-(\ref{2-degenerate}), we derive
\begin{eqnarray}
\Omega &=& \Omega_{0}
-\frac{\mu_e^4}{12 \pi^2}
+\frac{\Delta^2}{4G_D}
-\frac{\Lambda^4}{2\pi^2}
-\frac{\mu_{ub}^4}{12 \pi^2}
-\frac{\mu_{db}^4}{12 \pi^2} \nonumber \\
&-& 2 \int_{0}^{\Lambda} \frac{p^2 d p}{\pi^2}
\left(\sqrt{(p+ \bar{\mu})^2+\Delta^2}
+\sqrt{(p-\bar{\mu})^2+\Delta^2}\right)\nonumber \\
&-& 2\theta \left(\delta\mu-\Delta\right)
\int_{\mu^{-}}^{\mu^{+}}\frac{p^2 d p}{\pi^2}\Big(
\delta\mu-\sqrt{(p-\bar{\mu})^2+\Delta^2}
\Big), 
\label{pot0}
\end{eqnarray}
where $\mu^{\pm}\equiv \bar{\mu}\pm \sqrt{(\delta\mu)^2-\Delta^2}$. 
The appearance of the last term that contains the step function
$\theta(\delta\mu-\Delta)$ is remarkable. It can be traced back to 
the existence of two degenerate gapless modes in the quasiparticle
spectrum, whose energies change sign at the values of momenta
$p=\mu^{\pm}$, see Eq.~(\ref{2-degenerate}). In a different language, 
this sign change modifies the prescription for treating the quasiparticle
poles in the complex energy plane. This is exactly the reason why the
arguments of Ref.~\cite{enforced} regarding the conventional ``enforced"
pairing in color superconductors do not apply in the case of g2SC. 
The nonequal densities of pairing quarks, see Eq.~(\ref{n_dg-n_ur-T0}), 
is the formal consequence of this fact.

\subsection{Solving neutrality conditions}

Now, let us proceed to the analysis of the color and electrical charge
neutrality conditions in the model of dense quark matter at hand. At zero
temperature, the corresponding two independent conditions could be 
written in the form: $n_8=0$ and $n_Q+\half n_8=0$, i.e.,
\begin{eqnarray}
n_8 &=&\frac{4}{3}\int\frac{d^3{\bf p}}{(2\pi)^3}
\Bigg( - \frac{E-\bar\mu}{E_{\Delta}^-} \theta(E_{\Delta^-}^- )
+  \frac{E+\bar\mu}{E_{\Delta}^+} 
 - \theta(-E_{ub}^-)
 - \theta(-E_{db}^-) \Bigg) \nonumber \\
&\simeq& 
\frac{2}{9\pi^2}\left(2\bar\mu^3-\mu_{ub}^3-\mu_{db}^3\right)
+\frac{2 \bar\mu \Delta^2}{3\pi^2}\left(
\ln\frac{4(\Lambda^2-\bar\mu^2)}{\Delta^2}
-\frac{2\Lambda^2-\bar\mu^2}{\Lambda^2-\bar\mu^2} 
+\cdots\right) \nonumber \\
&& +  \frac{2\bar\mu}{3\pi^2} \theta\left(\delta\mu-\Delta \right)
\left(2\delta\mu \sqrt{(\delta\mu)^2-\Delta^2}- \Delta^2
\ln\frac{\delta\mu+\sqrt{(\delta\mu)^2-\Delta^2}}
{\delta\mu-\sqrt{(\delta\mu)^2-\Delta^2}}\right) \nonumber \\
&=&0,
\label{n8-T=0}
\end{eqnarray}
where the ellipsis denotes the terms of higher order in powers of 
$\Delta$, and
\begin{eqnarray}
n_Q+\half n_8 &=&\frac{\mu_e^3}{3\pi^2}
+2 \int\frac{d^3{\bf p}}{(2\pi)^3} \Bigg(
 - \theta(-E_{ub}^-) + \theta(-E_{\Delta^-}^-)
\Bigg) \nonumber \\
&=&\frac{\mu_e^3-\mu_{ub}^3}{3\pi^2} 
+\frac{2}{3\pi^2} \theta\left(\delta\mu-\Delta \right)
\sqrt{(\delta\mu)^2-\Delta^2}
\left(3\bar\mu^2+(\delta\mu)^2-\Delta^2 \right) \nonumber \\
&=& 0.
\label{nQ-T=0}
\end{eqnarray}
This last condition is particularly simple in the 2SC phase, 
$\Delta\geq \delta\mu$. In this case, the exact solution to this 
equation is available: 
\begin{equation}
\mu_e = \frac{3\mu-2\mu_{8}}{5}, \quad \mbox{for} \quad
\Delta\geq \delta\mu.
\end{equation}
The other equation, then, has the following approximate
solution:
\begin{eqnarray}
\mu_{8} &\simeq & \frac{45}{44}\left(\frac{4}{5}-6^{1/3}
-\frac{4^{1/3}}{3^{1/3}}\right)\mu +\cdots
\approx  0.0854\mu+\cdots ,
\end{eqnarray}
where the ellipsis again denotes the terms of higher order in powers of 
$\Delta/\mu$. As we see, the color chemical potential $\mu_{8}$ is very 
small compared to the other chemical potentials, $\mu_e$ and $\mu$. 
Of course, the corrections to the above solutions may become large 
with increasing the value of the gap $\Delta$. Therefore, strictly 
speaking, these formulas are reliable only in the weakly coupled regime
when $\Delta \ll \mu$. Our numerical results show, however, that 
$\mu_{8}$ remains also small at relatively large values of the gap. 
Moreover, the approximation $\mu_8\approx 0$ works well in the 
region $\Delta<\delta\mu$ and in a wide range of parameters in the 
model. In the following, we shall use this observation to derive many 
useful analytical results.

%%%%%%%%%%%%%%%%%%%%%%%%%%%%%%%%%%%%%%%%%%%%%%%%%%%%%%%%%%%%%%%%%%%%
\begin{figure}
\begin{center}
\epsfxsize=12cm
\epsffile[88 0 588 313]{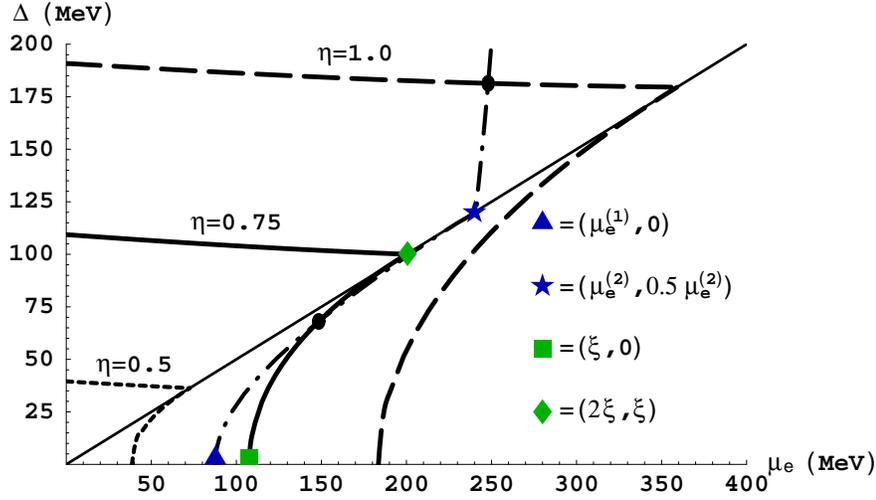}
\end{center}

\caption{The graphical representation of the solution to the charge 
neutrality conditions (thick dash-dotted line) and the solution to 
the gap equation for three different values of the diquark coupling 
constant (thick solid and dashed lines). The intersection points 
represent the solutions to both. The thin solid line divides two 
qualitatively different regions, $\Delta<\delta\mu$ and $\Delta>
\delta\mu$. The results are plotted for $\mu=400$ MeV and three 
values of diquark coupling constant $G_{D} = \eta G_S$ with $\eta=0.5$, 
$\eta=0.75$, and $\eta=1.0$.}
 
\label{gapneutral}
\end{figure}
%%%%%%%%%%%%%%%%%%%%%%%%%%%%%%%%%%%%%%%%%%%%%%%%%%%%%%%%%%%%%%%%%%%%

Now, let us consider the neutrality condition in the region $\Delta
\leq\delta\mu$. In our analytical treatment, we substitute $\mu_8= 0$. 
As we mentioned above, this is a very good approximation. In the case
of small values of $\Delta$ this should not be surprizing since the 
color charge vanishes in the normal phase when $\mu_8= 0$. The 
value of the electrical chemical potential is determined by solving 
the neutrality condition in Eq.~(\ref{nQ-T=0}). The corresponding 
result for $\mu_e^{(n_Q=0)}(\mu,\Delta)$ could be given by the inverse 
representation of the gap in terms of the potential $\mu_e$,
\begin{equation}
\Delta^{(n_Q=0)} = \sqrt{\frac{\mu_e^2}{4}-
\left(\frac{(\mu-2\mu_e/3)^3-\mu_e^3}
{6(\mu-\mu_e/6)^2}\right)^2} , 
\quad \mbox{for} \quad \Delta\leq\delta\mu.
\label{nQ-analitical}
\end{equation}
This analytical result is barely distinguishable from the numerical 
solution represented by the dash-dotted line in Fig.~\ref{gapneutral}.
Quark matter is positively charged to the left from the neutrality 
line, and it is negatively charged to the right.

From Eq.~(\ref{nQ-analitical}), we find an approximate point of the 
intersection of the neutrality line with the $\mu_e$-axis ($\Delta=0$).
This is denoted by a triangle in Fig.~\ref{gapneutral}. This point 
determines the value of the electrical chemical potential in the normal 
phase of neutral quark matter. Its approximate value is given by a 
solution to a cubic equation, obtained from Eq.~(\ref{nQ-analitical}) 
after substituting $\Delta=0$ on the left hand side. This value is 
$\mu_e^{(1)}\simeq 0.2196 \mu$. For example, at $\mu =400$ MeV, it 
is $87.84$ MeV which is in good agreement with the numerical result 
plotted in Fig.~\ref{gapneutral}.

Also, from Eq.~(\ref{nQ-analitical}), we determine the intersection of 
the neutrality line with the line $\Delta=\delta \mu$ that separates two
qualitatively different regions in the plane $(\mu_e,\Delta)$. This occurs 
at $\mu_e^{(2)}\simeq \frac{3}{5} \mu$ that corresponds to $\Delta^{(2)}
=\half \mu_e^{(2)}\simeq \frac{3}{10}\mu$. For the choice of the quark 
chemical potential $\mu =400$ MeV, used in Fig.~\ref{gapneutral}, the 
corresponding values are $\mu_e^{(2)} \simeq 240$ MeV and $\Delta^{(2)} 
\simeq 120$ MeV which are again in good agreement with the numerical 
result (cf., the location of the ``knee'' of the dash-dotted line in 
Fig.~\ref{gapneutral}, denoted by a star). 

Before concluding this subsection, we should emphasize that the 
charge densities in Eqs.~(\ref{n8-T=0}) and (\ref{nQ-T=0}) do not
depend explicitly on the diquark coupling constant. This is the 
property of the mean field approximation that would break after 
taking higher order corrections into account. It is reasonable
to assume that the corresponding dependence is weak even in the 
full theory. As we shall see below, this property is not shared by
the solution for the gap $\Delta$ which is very sensitive to changes 
in the coupling constant. This observation would have far reaching 
consequences below.

\subsection{Solving gap equation}

Now, we turn to the solution of the gap equation (\ref{gap-T}) at zero
temperature. In this case, the equation simplifies considerably. Its 
explicit form reads
\begin{eqnarray}
&&\hspace{-5mm} \Delta = 8G_D\int\frac{d^3{\bf p}}{(2\pi)^3} 
\Bigg( \frac{\Delta}{E_{\Delta}^-} \theta\left(E_{\Delta^-}^-\right)
+ \frac{\Delta}{E_{\Delta}^+}\Bigg) 
\simeq \frac{4G_D\Delta}{\pi^2}\Bigg[\Lambda^2-3\bar\mu^2\nonumber\\
&&\hspace{-5mm}+\left(\bar\mu^2-\half \Delta^2\right) 
\ln\frac{4(\Lambda^2-\bar\mu^2)}{\Delta^2}
+\frac{\Delta^2(\Lambda^4+3\Lambda^2\bar\mu^2-2\bar\mu^2)}
{2(\Lambda^2-\bar\mu^2)^2}+\cdots
\nonumber\\
&&\hspace{-5mm}-\theta\left(\delta\mu-\Delta \right)\left(
\delta\mu \sqrt{(\delta\mu)^2-\Delta^2}
+\left(\bar\mu^2-\half \Delta^2\right)
\ln\frac{\delta\mu+\sqrt{(\delta\mu)^2-\Delta^2}}
{\delta\mu-\sqrt{(\delta\mu)^2-\Delta^2}}\right)
\Bigg].
\label{gap-T=0}
\end{eqnarray}
As it should be, this equation always has the trivial solution 
$\Delta=0$. This corresponds to the normal phase of quark matter. 
If there is also a nontrivial solution $\Delta\neq 0$, the normal
phase is not necessarily the ground state. In fact, the true ground 
state corresponds to the solution that minimizes the value of the 
thermodynamic potential (i.e., it maximizes the pressure). Moreover, 
in the case of an additional requirement of neutrality, the 
minimization should be done along the neutrality line discussed 
in the preceding subsection.

For a moment, let us drop the condition of the charge neutrality
and consider all possible nontrivial solutions to the gap equation 
(\ref{gap-T=0}). We realize right away that there are two different 
branches of solutions in two regions, $\Delta<\delta\mu$ and
$\Delta>\delta\mu$, see Fig.~\ref{gapneutral}.

First, we start with the region $\Delta>\delta\mu$. This is the 
case of a relatively small mismatch between the Fermi momenta of the
pairing quarks. In particular, it describes the conventional 2SC 
phase of quark matter. From Eq.~(\ref{gap-T=0}), we derive the 
approximate analytical expression for the nontrivial solution,
\begin{equation}
\Delta \simeq  \xi(\bar\mu) ,
\quad \mbox{for} \quad \Delta>\delta\mu ,
\label{region1-T=0}
\end{equation}
where
\begin{equation}
\xi (\bar\mu) \equiv  2\sqrt{\Lambda^2-\bar\mu^2}\exp\left(
-\frac{\pi^2}{8 \bar\mu^2 G_D} 
+\frac{\Lambda^2-3\bar\mu^2}{2\bar\mu^2}
\right).
\label{def-xi}
\end{equation}
Since the average chemical potential $\bar\mu$ depends on $\mu_e$, 
see Eq.~(\ref{mu-bar}), the above solution for the gap is also a 
function of $\mu_e$. This dependence, however, is very weak because
$\mu_e $ enters $\bar\mu$ with a relatively small weight. This is 
also confirmed by our numerical results for three different values 
of the diquark coupling constant presented in Fig.~\ref{gapneutral}. 
With increasing $\mu_e$, the value of the gap decreases by less than
10\% along the whole upper branch of the solution. In contrast, the
dependence of the gap on the coupling constant is very sensitive.

In the region $\Delta<\delta\mu$, the mismatch parameter $\delta\mu$
is large. This is where the g2SC phase of quark matter appears. The
solution to the gap equation in this region is approximated well by 
the following expression:
\begin{equation}
\Delta \simeq \sqrt{\xi \left(\mu_e-\xi\right)},
\quad \mbox{for} \quad \Delta<\delta\mu,
\label{region2-T=0}
\end{equation}
where $\xi$ is the same weakly dependent function of $\mu_e$ that 
was defined in Eq.~(\ref{def-xi}). By making use of this analytical
expression, we see that the lower branch of the solution to the 
gap equation extends only over a finite range of chemical potentials,
$\mu_e^{(\rm min)} \leq \mu_e \leq \mu_e^{(\rm max)}$ (see 
Fig.~\ref{gapneutral}). The values of $\mu_e^{(\rm min)}$ and 
$\mu_e^{(\rm max)}$ are determined by the solutions to the equations
$\mu_e =\xi(\mu_e)$ and $\mu_e = 2\xi(\mu_e)$. The corresponding
graphical solutions are represented by the square and the diamond in 
Fig.~\ref{gapneutral}, respectively. At 
$\mu=400$ MeV and $\eta =0.75$, the corresponding approximate values 
are $\mu_e^{(\rm min)}\approx 107.9$ MeV and $\mu_e^{(\rm max)}\approx 
205.3$ MeV which are in reasonable agreement with numerical results 
in Fig.~\ref{gapneutral}. 

In summary of this subsection, we note that the general nontrivial 
solution to the gap equation consists of two different branches in two 
regions, $\Delta>\delta\mu$ and $\Delta<\delta\mu$. The upper is the 
main branch that extends down to $\mu_{e}=0$ (i.e., the case without a
mismatch in the Fermi surfaces of the up and down quarks). The lower 
branch appears only in a finite window of electrical chemical potentials.
It merges with the upper one at a point on the line $\Delta=\delta\mu$. 
In Fig.~\ref{gapneutral}, three solutions to the gap equation at $\mu=400$ 
MeV in three regimes with different coupling constants, corresponding to 
$\eta=0.5$, $\eta=0.75$, and $\eta=1.0$, are shown. The three curves 
have qualitatively the same shape, but differ by overall scale factors. 
As we shall see below, this difference in the overall scales has an important
physical consequence. In particular, at weak coupling ($\eta\lesssim 0.7$), 
the ground state of neutral quark matter is the normal phase, while 
neutral g2SC and 2SC phases are most stable at intermediate ($0.7 
\lesssim \eta \lesssim 0.8$) and strong ($\eta \gtrsim 0.8$) couplings, 
respectively. This is studied in detail in the following subsection.

\subsection{Ground state of neutral matter}
\label{gr-state} 

The ground state of neutral matter is determined by the location of 
the minimum of the effective potential as a function of the gap. It 
is understood that the minimization is performed after the neutrality 
condition is fulfilled. In different words, the consideration is 
restricted only to the neutrality (dash-dotted) line in 
Fig.~\ref{gapneutral}. Mathematically, the necessary condition for 
the minimum reads
\begin{equation}
\left(\frac{d \Omega}{d \Delta} \right)_{n_Q,n_8=0}
\equiv \frac{\partial \Omega}{\partial \Delta}  
+ \frac{\partial \Omega}{\partial \mu_e} 
\left(\frac{d \mu_e}{d \Delta} \right)_{n_Q,n_8=0}
+ \frac{\partial \Omega}{\partial \mu_8} 
\left(\frac{d \mu_8}{d \Delta} \right)_{n_Q,n_8=0} =0.
\label{necessary-cond}
\end{equation}
We notice that the last two terms with the derivatives taken along 
the neutrality line are proportional to the electrical and color charge 
densities. They become zero when the requirement of neutrality is 
enforced. Therefore, the above condition formally coincides with the 
gap equation studied in the preceding subsection. The important difference 
is that Eq.~(\ref{necessary-cond}) is defined only at the neutrality 
line in the $(\mu_e,\Delta)$ plane.

In order to satisfy the condition in Eq.~(\ref{necessary-cond}), it 
is very convenient to use a simple graphical construction, presented 
in Fig.~\ref{gapneutral}. The extremum values of the gap are determined
by the points of intersection of the neutrality (dash-dotted) line
with the solutions to the gap equation (i.e., either with one of the 
branches of the non-trivial solution, or with $\Delta=0$ line). 

At very weak coupling (e.g., $\eta=0.5$ case), both branches of the 
nontrivial solutions to the gap equation are squeezed into a small 
region around the origin in the $(\mu_e,\Delta)$ plane. As a result, 
the neutrality line does not intersect them. Therefore, the only extremum 
of the effective potential corresponds to the trivial solution $\Delta=0$. 
As is easy to see, this is the global minimum. Thus, the normal phase is 
the ground state of quark matter in this regime.

In the opposite case of very strong coupling (e.g., $\eta=1$ case), the
potential has two extrema. Of course, one of them corresponds to the 
normal phase ($\Delta=0$). This is a local maximum of the potential. 
The other extremum is a minimum that corresponds to the 2SC phase with 
$\Delta > \delta\mu$. Note that the types of the extrema could be inferred
from the expected topology of the effective potential after taking into 
account that this potential is bounded below when $\Delta\to \infty$.
Our direct calculation show that the 2SC phase is indeed the ground state. 
At large coupling, this is hardly a surprise.

\begin{figure}
\epsfxsize=12cm
\epsffile{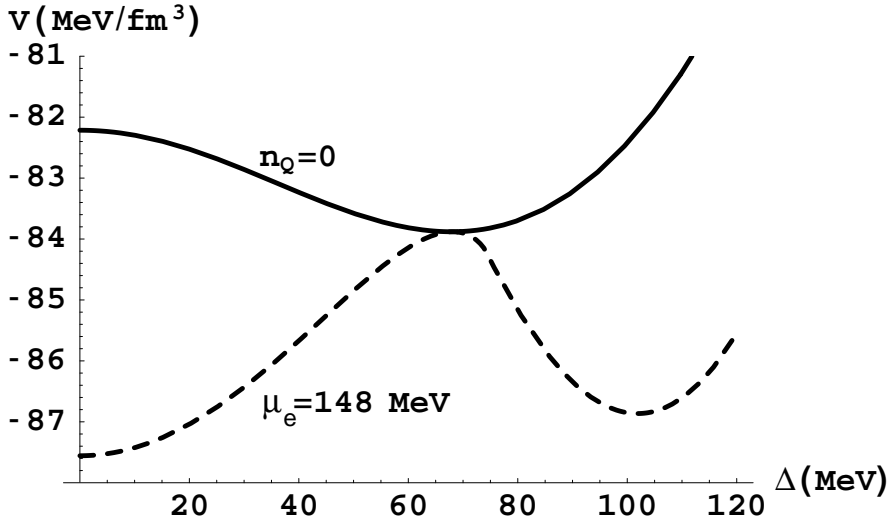}

\caption{The effective potential as a function of the diquark gap $\Delta$
calculated at a fixed value of the electrical chemical potential $\mu_e =
148.4$ MeV (dashed line), and the effective potential defined along the
neutrality line (solid line). The results are plotted for $\mu=400$ MeV
and $G_{D}=\eta G_{S}$ with $\eta=0.75$.}

\label{V2D}
\end{figure}

The regime of intermediate coupling (see, e.g., $\eta=0.75$ case) is 
most interesting. At first sight, it looks similar to the case of strong 
coupling because the potential has two extrema. One of them occurs 
at $\Delta=0$ and the other at $\Delta\neq 0$. The peculiar thing is 
that the nontrivial extremum corresponds to an intersection of the 
neutrality line with the {\em lower} branch (defined in the region 
$\Delta < \delta\mu$) of the nontrivial solutions to the gap equation. 
By making use of the same topological arguments as in the previous 
case, we conclude that the nontrivial solution defines the ground 
state of neutral quark matter. This can also be checked by a direct 
numerical calculation of the effective potential, see the solid line 
in Fig.~\ref{V2D}. As we shall discuss later in more detail, this 
ground state with $\Delta\neq 0$ is the g2SC, rather than ordinary 
2SC phase. 

%%%%%%%%%%%%%%%%%%%%%%%%%%%%%%%%%%%%%%%%%%%%%%%%%%%%%%%%%%%%%%%%%%%%
%%%%%%%%%%%%%%%%%%%%%%%%%%%%%%%%%%%%%%%%%%%%%%%%%%%%%%%%%%%%%%%%%%%%

At this point, it is relevant to make a few comments regarding the
early studies of the color superconducting quark matter at finite 
values of the electrical (or, equivalently, isospin) chemical potential 
\cite{bedaque}. There, the electrical chemical potential was treated as 
a free parameter, and the charge neutrality condition was not imposed.
Obviously, the 2SC phase is the ground state of quark matter at small 
values of the mismatch parameter $\delta\mu$ (which is identical with 
$\mu_e/2$ here). However, when the value of $\delta\mu$ increases and 
reaches a critical value, a first order phase transition happens. 
The approximate value of the critical electrical chemical potential 
is $\mu^{\rm cr}_{e}\simeq \sqrt{2} \Delta_{0}$ where $\Delta_{0}$ is 
the value of the gap at $\mu_{e} =0$ (strictly speaking, this estimate 
is derived in the approximation of weak coupling \cite{mag-crit}).
At higher values of $\delta\mu$, the normal phase is the new ground 
state of quark matter. In this picture, there is no room for the g2SC
phase. 

Now, how can one reconcile the appearance of the g2SC phase after
imposing the charge neutrality conditions, and its absence without
such conditions? To answer this, we refer again to Fig.~\ref{gapneutral}. 
In absence of the neutrality condition, one defines the potential 
$V(\Delta)$ by restricting $\Omega(\Delta,\mu_e)$ to a vertical line 
$\mu_e=\mbox{Const}$ for each fixed value of the parameter $\mu_e$. 
As a result, one ends up with a whole family of the effective potentials 
that are functions of $\Delta$. By changing the parameter $\mu_e$, 
one goes from one representative of the family to another. The extrema 
of the effective potentials are determined by the intersection points 
of $\mu_e=\mbox{Const}$ line with the line of solutions to the gap 
equation. 

Sweeping the values of $\mu_e$ from left to right in Fig.~\ref{gapneutral}, 
one easily understands the reason for the first order phase transition 
in the quark matter. At small $\mu_e$, the potential has two extrema 
at $\Delta=0$ and at $\Delta \neq 0$. Naturally, the latter is the ground 
state. At intermediate values of $\mu_e$ where the lower branch of the 
solution to the gap equation exists, $\mu_e^{\rm (min)}< \mu_e <
\mu_e^{\rm (max)}$, the potential has already three extrema that are given 
by the intersection of the line $\mu_e=\mbox{Const}$ with the $\mu_e$-axis 
(i.e., $\Delta=0$ solution) and with the two branches of the nontrivial 
solutions. Interestingly, the solution on the lower branch corresponds to 
a maximum of the potential. The same solution in the neutral matter was 
the global minimum of the properly defined potential, see Fig.~\ref{V2D}. 
The other two extrema are minima. It is the competition between them,
that results in a first order transition from the 2SC to the normal 
phase. 

It should be clear that this first order phase transition is unphysical 
under the requirement of local neutrality of quark matter. This is because 
the transition typically happens between two types of matter with different,
in general, {\em nonzero} charge densities (e.g., positively charged color
superconducting matter and negatively charged normal quark matter). It is
worth mentioning, however, that this first order phase transition can get
physical meaning in globally neutral mixed phases of quark matter
\cite{neutral_buballa,SHH}, and in some condensed matter systems 
where there is no analogue of the charge neutrality condition 
\cite{sarma}.

\subsection{Weak, intermediate and strong coupling regimes}
\label{w-i-s}

Let us discuss the rigorous meaning of the weak, intermediate and strong 
coupling regimes in the model at hand (as we shall see below, this could 
get somewhat ambiguous at finite temperature). We start from the weak 
regime first. By definition, the normal phase is the only neutral phase 
in this case. As should be clear, this happens when there is no intersection 
of the neutrality line with the two nontrivial branches of the solution 
to the gap equation. 

From Eq.~(\ref{nQ-analitical}) and the discussion following it, we know 
that the neutrality line intersects the $\mu_e$-axis at $\mu_e=\mu_e^{(1)}
\approx 0.22\mu$ (the corresponding point is marked by a triangle in 
Fig.~\ref{gapneutral}). The lower branch of the gap solution, on the other 
hand, starts at $\mu_e=\xi$, see Eq.~(\ref{region2-T=0}) and the point 
marked by a square in Fig.~\ref{gapneutral}. Therefore, these 
two lines do not intersect if $\xi < \mu_e^{(1)}$, and only the normal 
phase is consistent with the condition of neutrality, see the $\eta=0.5$ 
case in Fig.~\ref{gapneutral}. Now, we remind that while $\mu_e^{(1)}$ is 
essentially independent of the coupling constant, the value of $\xi$, 
for the definition see Eq.~(\ref{def-xi}), is very sensitive to the choice 
of the diquark coupling. Taking this into account, we derive the following 
approximate condition of the weakly coupled regime:
\begin{equation}
G_D < G_D^{(1)} \simeq 
\frac{\pi^2}{4\Lambda^2-12 \bar\mu_1^2+4\bar\mu_1^2
\ln\left[4\left(\Lambda^2-\bar\mu_1^2\right)/
\left(\mu_e^{(1)}\right)^2\right]},
\quad\mbox{``weak''},
\label{weak}
\end{equation}
with $\bar\mu_1 \simeq \mu-\frac{1}{6}\mu_e^{(1)}\approx 0.96\mu$ here.
At $\mu=400$ MeV, the above estimate gives $\eta \lesssim 0.684$. 

At larger coupling, the ground state of neutral matter is 
a broken phase. However, there are still two qualitatively different 
regimes depending on the location of the intersection of the neutrality 
line with the gap solution. If this occurs in the region $\Delta<\delta\mu$,
the g2SC phase is the ground state (e.g., $\eta=0.75$ case in 
Fig.~\ref{gapneutral}). Otherwise, the ground state is 2SC (e.g.,  
$\eta=1$ case in Fig.~\ref{gapneutral}). 

The criterion for the strong coupling regime reads $2\xi>\mu_e^{(2)}
=0.6 \mu$, i.e., $\xi>0.3\mu$. To derive this, we note that the value 
of the electrical chemical potential $\mu_e=0.6\mu$ corresponds to the 
intersection of the neutrality line with the boundary $\Delta=\delta\mu$ 
(this is marked by a star in Fig.~\ref{gapneutral}). If this value is 
smaller than $\mu_e=2\xi$ at which the gap solution intersects the boundary 
line (this is marked by a diamond in Fig.~\ref{gapneutral}), then the 
ground state is the 2SC phase. This is satisfied if the value of the 
coupling strength in the diquark channel is sufficiently strong, i.e.,   
\begin{equation}
G_D > G_D^{(2)} \simeq 
\frac{\pi^2}{4\Lambda^2-12 \bar\mu_2^2+4\bar\mu_2^2
\ln\left[4\left(\Lambda^2-\bar\mu_2^2\right)/
\left(\mu_e^{(2)}/2\right)^2\right]},
\quad\mbox{``strong''},
\label{strong}
\end{equation}
with $\bar\mu_2 \simeq \mu-\frac{1}{6}\mu_e^{(2)} = 0.9 \mu$ here. 
At $\mu=400$ MeV, this estimate gives $\eta \gtrsim 0.806$. 

Finally, the case of intermediate coupling strength is realized when 
$G_D^{(1)}\leq G_D \leq G_D^{(2)}$, or alternatively, $\mu_e^{(1)}
\simeq 0.22\mu \lesssim \xi \lesssim 0.3\mu \simeq \mu_e^{(2)}/2$. 
In this case, the ground state is the g2SC phase. We check that, at 
$\mu=400$ MeV, this is realized when $0.684 \lesssim \eta \lesssim 
0.806$. A representative solution in this regime (with $\eta=0.75$) 
is shown in Fig.~\ref{gapneutral} by the thick solid line.

We would like to mention that the critical values of the coupling constant 
$G_D^{(1)}$ and $G_D^{(2)}$ in Eqs.~(\ref{weak}) and (\ref{strong}) are 
functions of the quark chemical potential $\mu$, although the corresponding 
dependence is weak. Nevertheless, the above qualitative picture remains 
the same independent of the actual value of the quark chemical potential. 
Namely, there are always two critical values of the coupling constant that 
separate the three qualitatively different regimes.

\subsection{Quasiparticle spectra in 2SC and g2SC phase}

In the previous subsections, we established that the stable ground 
state in neutral quark matter is the g2SC phase, provided the 
coupling constant is neither too weak nor too strong ($G_D^{(1)} 
\leq G_D \leq G_D^{(2)}$). However, these calculations alone may not 
be sufficient to understand the physical properties of such a phase. 
To compensate this limitation, we find it very useful to study the 
nature of the fermionic quasiparticles in the g2SC phase.

It is instructive to start with the case of the ordinary 2SC phase 
($\Delta>\delta\mu$). All four fermionic quasiparticles that result 
from the Cooper pairing of the red and green quarks are gaped (note 
that, with the conventional choice of the gap pointing in the anti-blue 
direction in color space, the spectra of the blue up and blue down quarks 
are not affected by the pairing dynamics). 

\begin{figure}
\hbox{
\epsfxsize=6cm
\epsffile{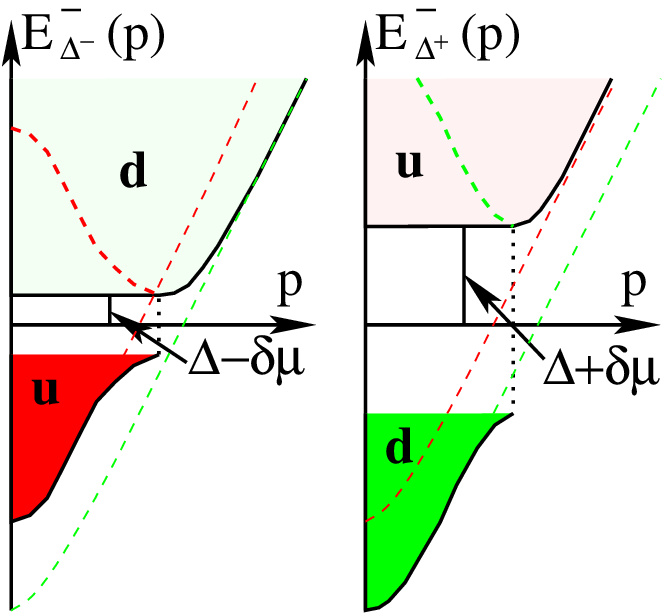}
\epsfxsize=6.5cm
\epsffile{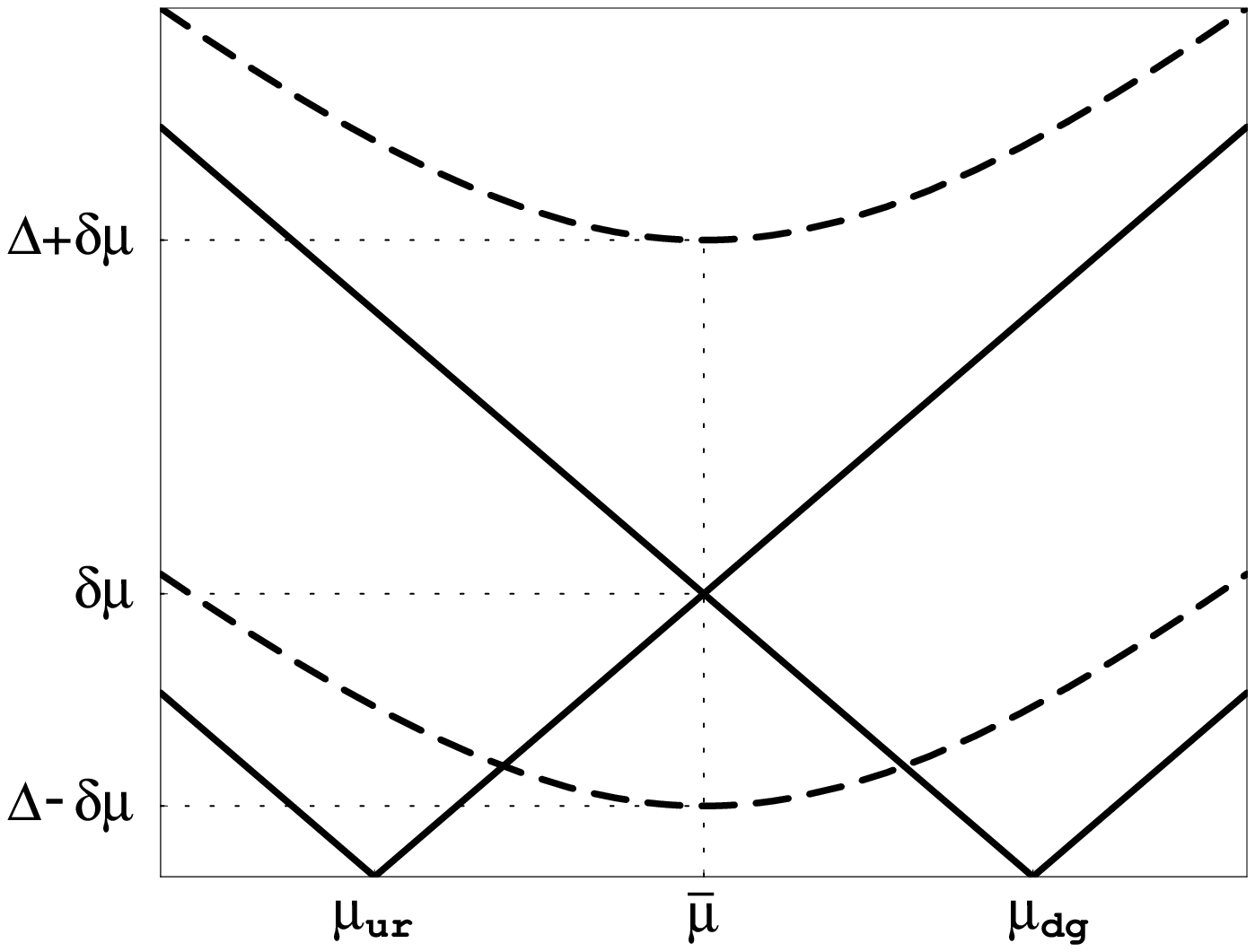}
}
\caption{The schematic modification of the red and green quark 
dispersion relations in the 2SC phase (left panel), and the 
resulting quasiparticle dispersion relations at low energies 
(right panel).}

\label{2sc-quasi}
\end{figure}

The corresponding spectra are 
shown schematically in Fig.~\ref{2sc-quasi}. The left panel gives a 
qualitative picture of the dispersion relations of the pairing quarks, 
see Eq.~(\ref{2-degenerate}). The dark (light) shaded regions 
below (above) $E=0$ line represent the filled (empty) quark states.
By combining the lines lying above $E=0$ with the reflections of the 
lines lying below $E=0$, one obtains the complete dispersion relations 
of the fermionic quasiparticles in the 2SC phase. The low energy slices 
of these relations are shown in the right panel by dashed lines. For 
comparison, the dispersion relations of quasiparticles in free theory 
are also shown in the same panel by solid lines.

%%%%%%%%%%%%%%%%%%%%%%%%%%%%%%%%%%%%%%%%%%%%%%%%%%%%%%%%%%%%%%%%%%%%
%%%%%%%%%%%%%%%%%%%%%%%%%%%%%%%%%%%%%%%%%%%%%%%%%%%%%%%%%%%%%%%%%%%%

It is interesting to observe how the nature of quasiparticles changes 
gradually when one goes from the region of small momenta to the region 
of large momenta. From Fig.~\ref{2sc-quasi}, we see that the 
quasiparticles with the larger (smaller) gap are nearly identical with 
the down (up) quark holes at small momenta, $p\ll\bar\mu$, and resemble 
the up (down) quarks at large momenta, $p\gg\bar\mu$. Around the middle 
point, $p\simeq\bar\mu$, the quark spectrum is strongly modified by the 
appearance of the gap. Thus, the quasiparticles represent presumably a 
mixture of different types of quarks and holes.

\begin{figure}
\hbox{
\epsfxsize=6cm
\epsffile{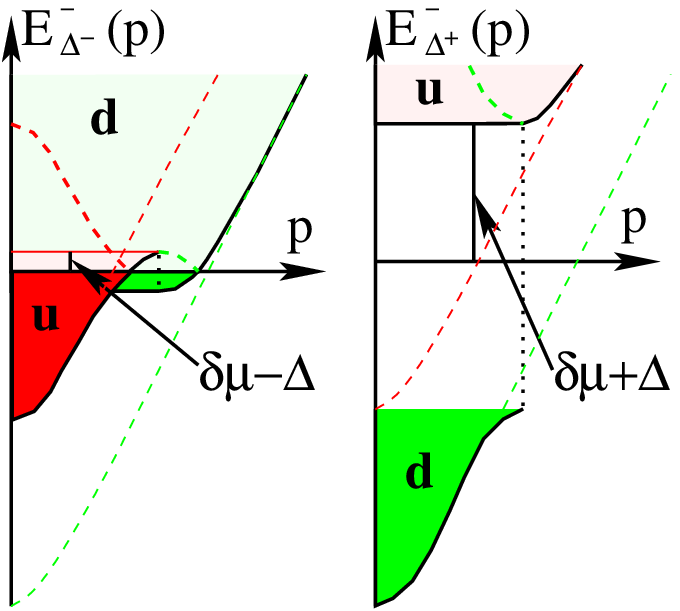}
\epsfxsize=6.5cm
\epsffile{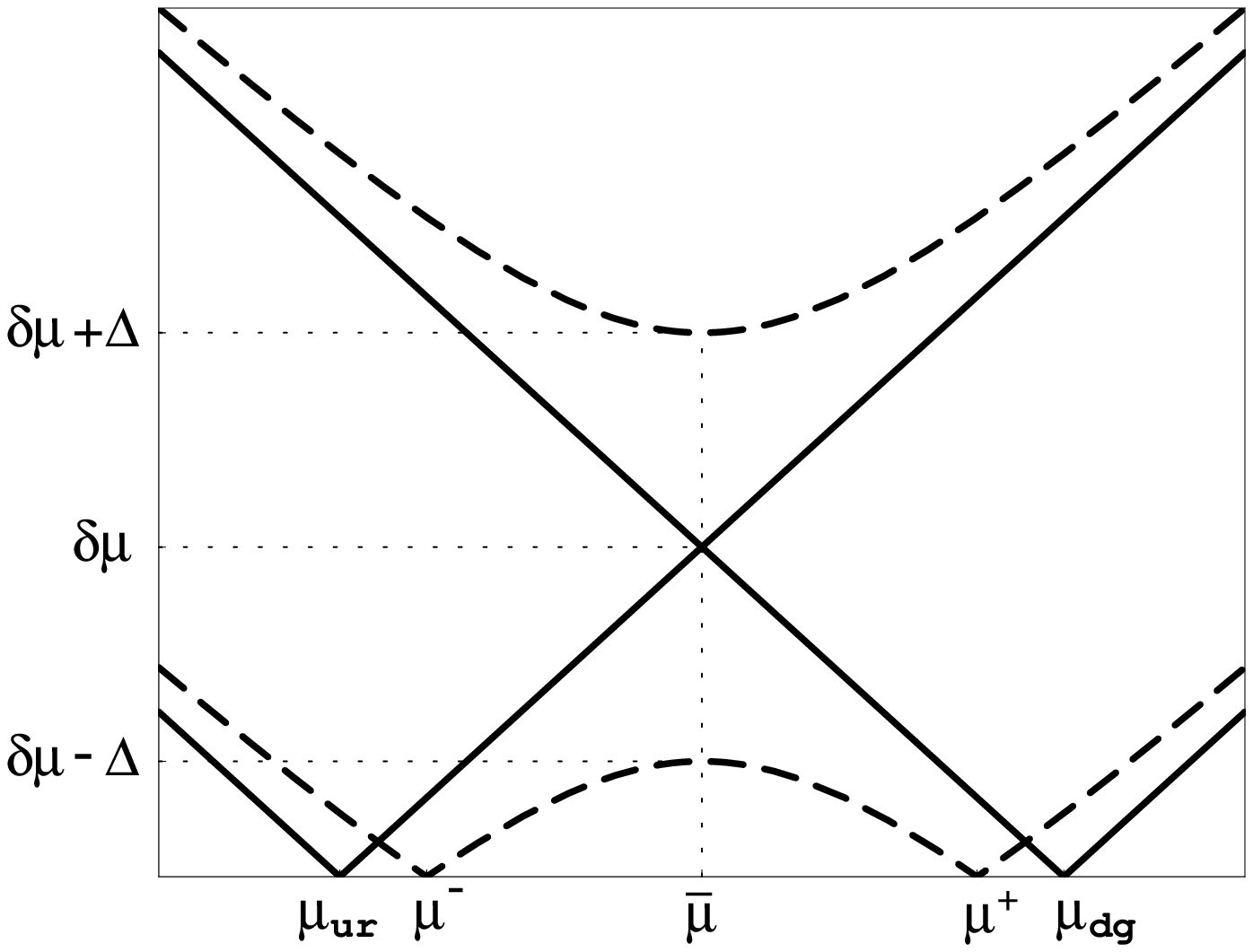}
}
\caption{The schematic modification of the red and green quark 
dispersion relations in the g2SC phase (left panel), and the 
resulting quasiparticle dispersion relations at low energies 
(right panel).}

\label{g2sc-quasi}
\end{figure}

Now, let us turn to the quasiparticle spectra in the g2SC phase. 
Graphically, these are shown in Fig.~\ref{g2sc-quasi} which has 
the same layout as Fig.~\ref{2sc-quasi}. The left panel shows
the qualitative dispersion relations of quarks in the case of 
$\Delta<\delta\mu$, see Eq.~(\ref{2-degenerate}), while the right 
panel gives the low energy slice of the corresponding quasiparticle 
dispersion relations (dashed lines).

As one would expect, far outside the pairing region, $p\simeq\bar\mu$, 
the quasiparticle dispersion relations are similar to those in the 2SC 
phase. Also, around $p\simeq\bar\mu$, the gaped (double-degenerate) 
quasiparticle resembles the (double-degenerate) quasiparticle with 
larger value of the gap in the 2SC phase. This is the quasiparticle 
that interpolates between the down quark holes at small momenta, 
$p\ll\bar\mu$, and the up quarks at large momenta, $p\gg\bar\mu$. 
The other (double-degenerate) quasiparticle is gapless. The energy
of this latter vanishes at two values of momenta $p=\mu^{-}$ and 
$p=\mu^{+}$ where $\mu^{\pm}\equiv \bar\mu\pm\sqrt{(\delta\mu)^2
-\Delta^2}$.

The most remarkable property of the quasiparticle spectra in the 
g2SC phase, see the right panel in Fig.~\ref{g2sc-quasi}, is that 
the low energy excitations ($E\ll \delta\mu-\Delta$) are very similar
to those in the normal phase represented by solid lines. The only 
difference is that the values of the chemical potentials of the up
and down quarks $\mu_{ur}=\mu_{ug}$ and $\mu_{dg}=\mu_{dr}$ are replaced 
by the values $\mu^{-}$ and $\mu^{+}$, respectively. This observation 
suggests, in particular, that the low energy (large distance scale) 
properties of the g2SC phase should look similar to those in the 
normal phase. In the following subsection, we use this argument to 
justify that the Meissner effect might be absent in the g2SC phase.

%%%%%%%%%%%%%%%%%%%%%%%%%%%%%%%%%%%%%%%%%%%%%%%%%%%%%%%%%%%%%%%%%%%%
%%%%%%%%%%%%%%%%%%%%%%%%%%%%%%%%%%%%%%%%%%%%%%%%%%%%%%%%%%%%%%%%%%%%

\subsection{Gluon properties in g2SC phase}

The main purpose for studying the NJL model in Eq.~(\ref{lagr}) is
to understand the properties of dense quark matter under conditions
that may exist inside compact stars. Indeed, this model should be adequate 
for getting an insight into the pairing dynamics of quarks and into 
some thermodynamic properties of quark matter. However, the NJL model 
lacks gluons. As reflection of this, it possesses the global instead 
of gauged color symmetry. The consequence is that there appear five 
Nambu-Goldstone (NG) bosons in a ground state of the model when the 
color symmetry is broken. In QCD, there is no room for such NG bosons. 

Perhaps, the NJL model can be thought of as the low energy theory 
of QCD in which the gluons, as independent degrees of freedom, are 
integrated out. The gluons could be reintroduced back by gauging the 
color symmetry in the Lagrangian density in Eq.~(\ref{lagr}), providing
a semirigorous framework for studying the effect of the Cooper pairing 
on the physical properties of gluons. 

Gluon properties in the 2SC phase of dense quark matter without the
mismatch of the Fermi momenta of the up and down quarks have been 
studied in detail \cite{R-2sc,RS-2sc}. A similar quantitative analysis 
could also be performed in the case of a nonzero mismatch parameter 
$\delta\mu$. Technically, this is not easy, and it is left for a 
future study. Here we argue, nevertheless, that some basic properties 
of gluons could be understood even without such an explicit calculation. 

At large densities, gluons become quasiparticles with rather 
complicated dispersion relations. Typically, the only propagating 
(i.e., weakly damped) modes of gluons in dense medium are plasmons 
(see, however, Ref.~\cite{npb700} where other interesting modes of 
dense quark matter are discussed). As in the case of metals, the 
minimum value of the frequency of plasmons is called the plasma 
frequency. Since its value in dense quark matter is of order 
$\sqrt{\alpha_s}\mu$, the plasmons could be viewed as ``high energy'' 
modes that exist not only in the normal phase, but also in the 
2SC and g2SC phases. The reason is that the properties of these high
energy modes should not change much by the modification of the low 
energy quasiparticle spectra in broken phases. 

It is clear that the gluon self-energy carries information not only 
about the propagating modes, but also about highly damped modes and 
various screening effects. The Meissner effect and the Debye screening
are among them. It is important to mention that these types of large 
distance scale properties are determined to large extent by the low 
energy excitations. Therefore, they are very sensitive to the 
(dis-)appearance of gaps in quasiparticle spectra. 

First, let us discuss the Debye screening which is the property of
the gluons of the electric type (i.e., spatially longitudinal gluons). 
In the case of the 2SC phase, this is studied in Ref.~\cite{R-2sc}. 
It is shown that the three gluons of the unbroken $SU(2)_c$, unlike 
the other five gluons of the original $SU(3)_c$, do {\it not} 
experience the Debye screening in the far infrared region. This is 
the consequence of the absence of gapless quasiparticles in the 2SC 
ground state that interact with the gluons of the unbroken $SU(2)_c$. 
Now, this property is no longer valid in the gapless phase studied 
here. One should expect, therefore, that all gluons of the original 
$SU(3)_c$ are subject to the Debye screening. Of course, the values 
of the Debye masses for different gluons need not be the same.

A similar line of arguments may also be given in relation to the 
Meissner effect that affects five magnetic gluons (i.e., spatially 
transverse gluons). In this case, it is the appearance of a nonzero 
gap that leads usually to Meissner masses for the gluons 
\cite{R-2sc,RS-2sc}. We remind that the three gluons of the unbroken 
$SU(2)_c$ do not experience the Meissner effect. It is reasonable to
suggest that, even with a nonzero mismatch parameter, five magnetic
gluons in the 2SC phase ($\Delta >\delta\mu$) should still experience
the Meissner effect. Indeed, all those quasiparticles that were gaped 
at $\delta\mu=0$ still remain gaped. 

In the case of the g2SC phase ($\Delta <\delta\mu$), there appear 
gapless quasiparticles in the low energy sector of the theory, see 
Fig.~\ref{g2sc-quasi}. If the value of the quantity $\delta\mu - \Delta$ 
is not vanishingly small, the low energy excitations ($E\lesssim 
\delta\mu - \Delta$) in the g2SC phase look similar to the excitations 
in the normal phase in which the values of the chemical potentials are 
shifted slightly. Moreover, even the values of the density of states at 
the corresponding ``effective'' Fermi surfaces, i.e., at $p=\mu^{-}$ and 
$p=\mu^{+}$, remain similar. By making use of the dispersion relations 
in Eq.~(\ref{2-degenerate}), we derive
\begin{eqnarray}
\left. \frac{dN}{dE}\right|_{p=\mu^{-}}
=\frac{2\delta\mu (\mu^-)^2}
{\sqrt{(\delta\mu)^2-\Delta^2}}, &\qquad &
\left. \frac{dN}{dE}\right|_{p=\mu^{+}} 
=\frac{2\delta\mu (\mu^+)^2}
{\sqrt{(\delta\mu)^2-\Delta^2}}.
\end{eqnarray}
Taking this into account, we may conclude that the large distance 
properties of the magnetic gluons in the g2SC phase should be the 
same as in the normal phase. This suggests, in particular, that 
there is no color Meissner effect in such a phase. This would be 
very unusual because the diquark gap is nonzero, and the 
Anderson-Higgs mechanism is expected to occur. Sometimes, this 
latter is identified with the Meissner effect. We emphasize, however, 
that the two effects are not the same. Of course, the Anderson-Higgs 
mechanism is realized in the g2SC phase in the following sense. The 
five NG bosons, related to the breaking of the global color symmetry, 
are unphysical in the g2SC phase. In other words, they are ``eaten'' 
by the gluons. This conclusion can be reached by using the gauge freedom 
in QCD that allows to get rid of the NG bosons (the corresponding gauge 
is called unitary). 

Here, it is also instructive to mention that, in contrast to models with 
unbroken Lorentz symmetry, the gauge bosons in dense (or hot) medium have 
propagating (massive) longitudinal modes even in the normal phase. Thus, 
the intuitive picture, being popular in particle physics, that connects 
the Anderson-Higgs mechanism with the appearance of nonzero masses for 
gauge bosons is not directly generalized to dense (or hot) medium. In the
case of the g2SC phase, it might not be adequate even when applied to the 
appearance of the Meissner mass for magnetic gluons.

\subsection{Spin-1 condensation}

The fact that the low energy fermionic quasiparticles in the g2SC 
phase look nearly the same as in the normal phase has yet another 
consequence. As we discuss below, there should appear two additional 
spin-1 condensates on top of the g2SC ground state. In its turn, this 
should lead to the opening of small gaps around the points $p=\mu^{-}$ 
and $p=\mu^{+}$. Thus, strictly speaking, there will be no gapless 
excitations in the quasiparticle spectrum at zero temperature.

Our argument here is based on the following simple observation. Around 
the effective Fermi momenta $p=\mu^{-}$ and $p=\mu^{+}$, the gapless 
quasiparticles are double degenerate. This is the reflection of the 
degeneracy with respect to the red and green colors. Then, by taking
into account the high degeneracy of states at $p=\mu^{-}$ and $p=\mu^{+}$
and the attraction in the color antisymmetric channel, we conclude that 
the Cooper instability should develop in the g2SC ground state around 
both points $p=\mu^{-}$ and $p=\mu^{+}$. This instability is removed 
spontaneously by the appearance of two ``secondary'' condensates of 
Cooper pairs. In view of the Pauli principle, these are spin-1 
condensates of the so-called polar phase. The spatial rotation symmetry 
is broken in the presence of such a condensation \cite{spin-1}.  

Since spin-1 condensates are much weaker than the spin-0 condensates,
this additional condensations add a fine structure on the background
of the g2SC phase. In practical terms, it means that many thermodynamic
quantities, such as the energy density, the pressure and the particle 
densities should change very little. At the same time, other properties, 
such as the specific heat, the magnetic field penetration depth and 
various transport coefficients could be modified considerably by the 
appearance of the spin-1 condensates. For example, the presence of 
these condensates will result in the color as well as electromagnetic 
Meissner effect at sufficiently low temperatures \cite{spin-1-Meissner}. 

Before concluding this subsection, we would like to mention that an
additional spin-1 condensation of the blue up and down quarks is not 
expected in the model at hand. Even if there is an attraction in a 
flavor antisymmetric channel (e.g., provided by instantons), the 
Fermi momenta of the blue up and down quarks are badly mismatched. This 
is the general outcome of enforcing the $\beta$-equilibrium condition 
and the local neutrality in nonstrange quark matter.

\section{Gapless 2SC at finite temperature}
\label{finite-T}

In this section, we study the properties of the g2SC phase at finite
temperature. As we shall see, the interplay of the neutrality condition
and the solution to the gap equation produces many unusual properties
of quark matter that have no analogue in well known systems. One of 
our most intriguing observation will be that the ratio of the critical 
temperature to the value of the gap at zero temperature is not a universal 
number in the g2SC phase. In fact, depending on the coupling constant, 
this ratio could be arbitrarily large. The other striking property is
related to a nonmonotonic dependence of the order parameter on temperature. 
In the most extreme case, for example, a nonzero gap can appear at finite 
temperature even it was exactly zero at $T=0$.

In a superconducting system, when one increases the temperature at a given 
chemical potential, thermal motion will eventually break up the quark 
Cooper pairs. In the weakly interacting Bardeen-Copper-Schrieffer (BCS) 
theory, the transition between the superconducting and normal phases is 
usually of second order. The ratio of the critical temperature 
$T_c^{\rm BCS}$ to the zero temperature value of the gap 
$\Delta_0^{\rm BCS}$ is a universal value \cite{ratio-in-BCS}
\begin{eqnarray}
r_{\rm BCS}=\frac{T_c^{\rm BCS}}{\Delta_0^{\rm BCS}} =
\frac{{\rm e}^{\gamma_E}}{\pi} \approx 0.567,
\label{r_BCS}
\end{eqnarray}
where $\gamma_E \approx 0.577$ is the Euler constant.
In the conventional 2SC phase of quark matter with equal densities of
the up and down quarks, the ratio of the critical temperature to the 
zero temperature value of the gap is also the same as in the BCS theory 
\cite{PR-sp1}. In the spin-0 color flavor locked phase as well as 
in the spin-1 color spin locked phase, on the other hand, this ratio
is larger than BCS ratio by the factors $2^{1/3}$ and $2^{2/3}$, 
respectively. These deviations are related directly to the presence 
of two different types of quasiparticles with nonequal gaps 
\cite{ratio-in-CFL}. 

This commonly accepted picture of the finite temperature effects in 
superconducting phases changes drastically in the case of dense quark 
matter when the $\beta$-equilibrium and the local neutrality conditions 
are enforced. Below, we study this in detail.

\subsection{Gap equation and charge neutrality condition}

Here we consider the gap equation and charge neutrality conditions in 
the g2SC phase at finite temperature. Because of additional technical 
complications appearing at finite temperature, most of the results that 
follow will be obtained by numerical computation. It is important to 
keep in mind, however, that our approach remains conceptually the 
same as that at zero temperature.
  
Before studying the gap equation (\ref{gap-T}), we recall that the right 
hand side of this equation is a function of the three chemical potentials 
$\mu$, $\mu_e$ and $\mu_8$. The values of $\mu_e$ and $\mu_8$ are not the 
free parameters in neutral quark matter. They are determined by satisfying 
the two conditions in Eq.~(\ref{Q=0}). In our analysis, on the other hand, 
it is very convenient to enforce only the color charge neutrality by 
choosing the function $\mu_8(\mu,\mu_e,\Delta)$ properly. As for the 
electrical chemical potential $\mu_e$, it is treated as an independent 
parameter at intermediate stages of calculation. Only at the very end its 
value is adjusted to make the quark matter also electrically neutral.

As in the case of zero temperature, the resulting color chemical potential 
$\mu_8$ is much smaller than the other two chemical potentials in neutral 
quark matter. This, of course, is not surprizing. Small nonzero values of 
$\mu_8$ [typically, $\mu_8 \sim \Delta^2/\mu$] are required only because 
an induced color charge of the diquark condensate should be compensated.
Therefore, the smallness of $\mu_8$ is protected by the smallness of the
order parameter. The numerical calculations show, in fact, that even 
the approximation $\mu_8=0$ does not modify considerably the exact solutions 
to the gap equation and to the electrical charge neutrality condition. In 
practice, we either tabulate the function $\mu_8(\mu,\mu_e,\Delta)$ for a 
given set of parameters, or determine it numerically in the vicinity of 
the ground state solution.

\begin{figure}
\epsfxsize=12cm
\epsffile{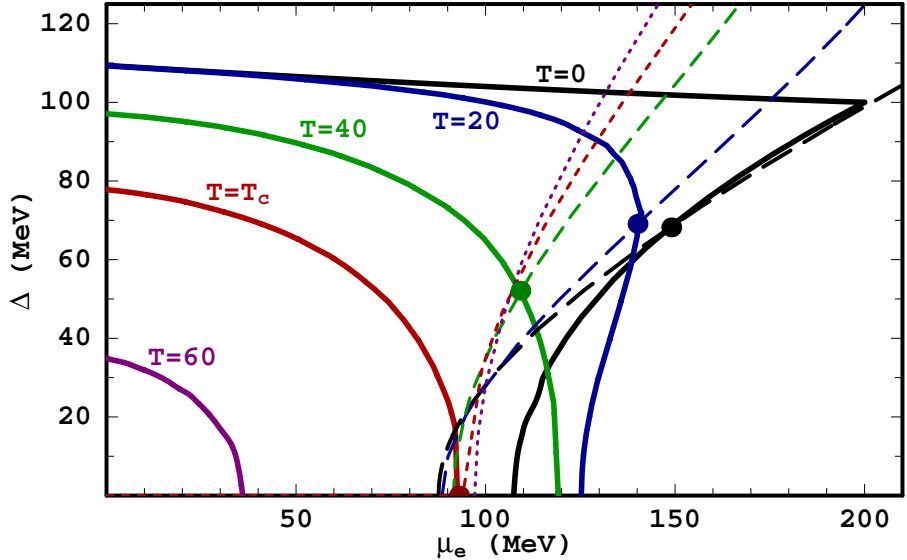}

\caption{The solutions to the gap equation (solid lines) and the 
neutrality condition (dashed lines) calculated for several values of 
temperature. The results are plotted for $\mu=400$ MeV and the 
diquark coupling constant $G_D=\eta G_S$ with $\eta=0.75$. In this 
case, $T_c\approx 50.2$ MeV.}
\label{gap-vs-mue-T.eps}
\end{figure}

The solutions to the gap equation (\ref{gap-T}) for several different 
values of temperature ($T=0$, $20$, $40$, $50.2$ and $60$ MeV) are shown 
graphically in Fig.~\ref{gap-vs-mue-T.eps} (solid lines). The values of 
the temperature are marked along the curves. The results are plotted in 
the ($\mu_e$,$\Delta$)-plane. In computation we kept the quark chemical 
potential fixed, $\mu=400$ MeV, and used the diquark coupling constant 
$G_D=\eta G_S$ with $\eta=0.75$. This is the same choice that we used at
zero temperature in Fig.~\ref{gapneutral}.
   
%%%%%%%%%%%%%%%%%%%%%%%%%%%%%%%%%%%%%%%%%%%%%%%%%%%%%%%%%%%%%%%%%%%%
%%%%%%%%%%%%%%%%%%%%%%%%%%%%%%%%%%%%%%%%%%%%%%%%%%%%%%%%%%%%%%%%%%%%

Now let us briefly discuss the temperature dependence of the solution. 
It is not very surprizing that the shape of the graphical solution is 
smoothed with increasing the temperature. The same applies to the 
disappearance of the double-branch structure of the solution at a 
finite value of the temperature, $T_s \simeq 30$ MeV. Indeed, in a 
model that treats the electrical chemical potential as a free parameter, 
this value of the temperature $T_{s}$ marks the expected switch of two 
regimes. Namely, while the phase transition controlled by the $\mu_e$ 
parameter (i.e., neutrality is not required) is a first order phase 
transition at $T<T_s$, it becomes a second order phase transition at 
$T>T_s$. 

We also observe that the value of the electrical chemical potential 
at the point where the solution to the gap equation intersects with 
the $\mu_e$-axis has a nonmonotonic dependence on the temperature. With
increasing temperature, this value increases first and, after reaching 
some maximum value, starts to decrease and goes down to zero eventually.
When the interplay with the neutrality condition is taken into account
later, this simple property of the solution would produce rather unusual 
physical results.

We note that the temperature dependence of the solutions to the gap 
equation is very similar to the dependence found by Sarma in a  
solid state physics analogue of the g2SC phase \cite{sarma}. Of 
course, there was no analogue of the neutrality condition in the
system studied in Ref.~\cite{sarma}. Therefore, the mentioned similarity 
does not extend to the complete analysis of the g2SC phase that 
follows.

In Fig.~\ref{gap-vs-mue-T.eps}, we also show the neutrality lines (dashed
lines) for the same values of temperature ($T=0$, $20$, $40$, $50.2$ and 
$60$ MeV). The convention is that the lengths of the dashes decreases with 
increasing the value of the temperature. As we see, with increasing 
temperature, the neutrality line gets steeper while its intersection 
point with the $\mu_e$-axis moves towards larger values of $\mu_e$. The 
points of intersection of these neutrality lines with the lines of 
solutions to the gap equation, when they exist, are shown as well. 

\begin{figure}
\epsfxsize=12cm
\epsffile{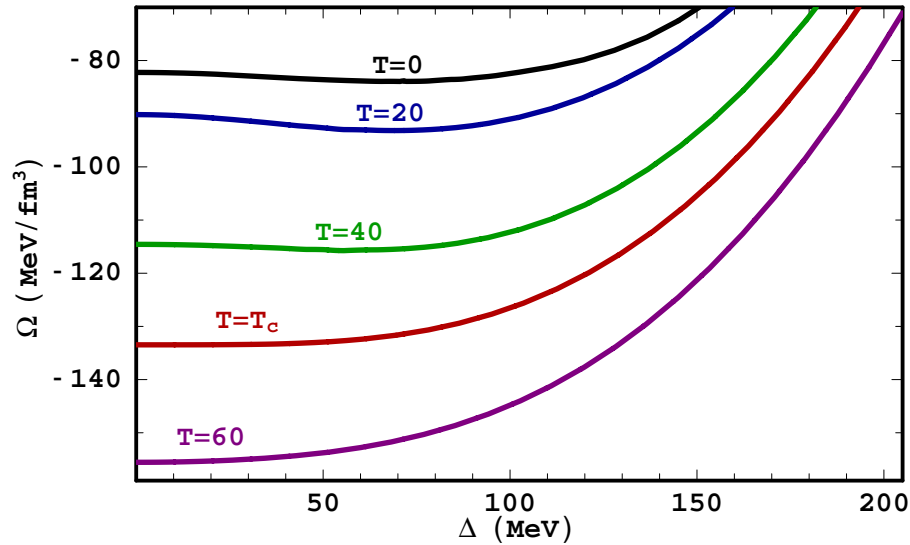}

\caption{The thermodynamic potential of neutral quark matter as a 
function of the diquark gap $\Delta$ calculated for several values of 
temperature.} 
\label{potdd}
\end{figure}

As in the case of zero temperature, see Subsec.~\ref{gr-state} and 
Fig.~\ref{V2D}, we need to show that the points of
intersections of the gap solutions with the corresponding neutrality 
lines in Fig.~\ref{gap-vs-mue-T.eps} represent the ground state of the 
neutral quark matter. To this end, we calculate the dependence of the 
thermodynamic potential on the value of the gap $\Delta$. Since the 
charge neutrality condition is satisfied only along the neutrality 
line, we restrict the thermodynamic potential only to this line. The 
numerical results for several values of the temperature are shown in 
Fig.~\ref{potdd}.

%%%%%%%%%%%%%%%%%%%%%%%%%%%%%%%%%%%%%%%%%%%%%%%%%%%%%%%%%%%%%%%%%%%%
%%%%%%%%%%%%%%%%%%%%%%%%%%%%%%%%%%%%%%%%%%%%%%%%%%%%%%%%%%%%%%%%%%%%

The results at $\mu=400$ MeV and $\eta=0.75$ show that, for any 
$T<T_c\approx 50.2$ MeV, the thermodynamic potential 
has the global minimum away from the origin, meaning that the 
corresponding ground state develops a nonzero expectation value of 
$\Delta(T)$. Needless to say that this is the same value that one 
extracts from the location of the geometrical construction in 
Fig.~\ref{gap-vs-mue-T.eps}. The expectation value $\Delta(T)$
disappears gradually when the temperature approaches $T_{c}$ from
below. This is an indication of the second order phase transition. At 
temperature higher than $T_c$, the ground state of neutral quark 
matter is the normal phase with $\Delta(T)=0$.

\subsection{Temperature dependence of the gap}

The temperature dependence of the gap is obtained by numerical solution
of the gap equation (\ref{gap-T}) together with the two charge neutrality 
conditions in Eq.~(\ref{Q=0}). Of course, this is equivalent to the 
geometrical construction used in Fig.~\ref{gap-vs-mue-T.eps}, where the
intersection points of two types of lines determine the values of the
gaps in the ground state. 

\begin{figure}
\epsfxsize=12cm
\epsffile{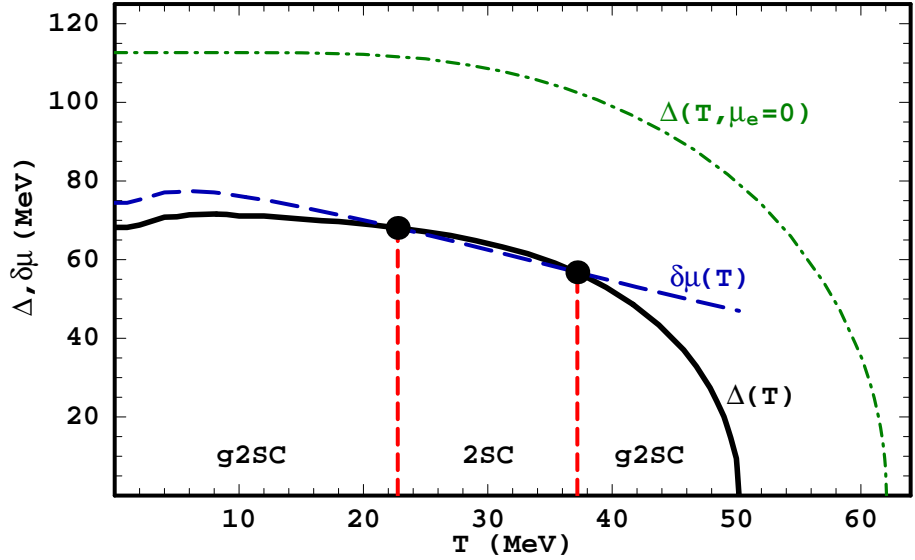}

\caption{The temperature dependence of the diquark gap (solid line) 
and the value of $\delta\mu\equiv \mu_e/2$ (dashed line) in neutral 
quark matter. For comparison, the diquark gap in the model with 
$\mu_e=0$ and $\mu_8=0$ is also shown (dash-dotted line). The results 
are plotted for $\mu=400$ MeV and $\eta=0.75$.}

\label{gap-mue0-vs-T.eps}
\end{figure}

The typical results for the default choice of parameters $\mu=400$ MeV 
and $\eta=0.75$ are shown in Fig.~\ref{gap-mue0-vs-T.eps}. Both the 
values of the diquark gap (solid line) and the mismatch parameter 
$\delta\mu=\mu_e/2$ (dashed line) are plotted. One very unusual property 
of the shown temperature dependence of the gap is the nonmonotonic behavior.
Only at sufficiently high temperatures, the gap is a decreasing function. 
In the low temperature region, $T\lesssim 10$ MeV, however, it increases 
with temperature. For comparison, in the same figure, the diquark gap 
in the model with $\mu_e=0$ and $\mu_8=0$ is also shown (dash-dotted 
line). This latter has the standard BCS shape.

%%%%%%%%%%%%%%%%%%%%%%%%%%%%%%%%%%%%%%%%%%%%%%%%%%%%%%%%%%%%%%%%%%%%
%%%%%%%%%%%%%%%%%%%%%%%%%%%%%%%%%%%%%%%%%%%%%%%%%%%%%%%%%%%%%%%%%%%%

Another interesting thing regarding the temperature dependences in
Fig.~\ref{gap-mue0-vs-T.eps} appears in the intermediate temperature 
region, $22.5 \lesssim T \lesssim 37$ MeV. By comparing the values of 
$\Delta(T)$ and $\delta\mu$ in this region, we see that the g2SC phase 
is replaced by a ``transitional'' 2SC phase there. Indeed, the 
energy spectrum of the
quasiparticles even at finite temperature is determined by the same 
relations in Eqs.~(\ref{disp-ub}) and (\ref{2-degenerate}) that we 
used at zero temperature. When $\Delta > \delta \mu$, the modes 
determined by Eq.~(\ref{2-degenerate}) are gaped. Then, according 
to our standard classification, the ground state is the 2SC phase.

It is fair to say, of course, that the qualitative difference of the 
g2SC and 2SC phases is not so striking at finite temperature as it 
is at zero temperature. This difference is particularly negligible 
in the region of interest where temperatures $22.5 \lesssim T 
\lesssim 37$ MeV are considerably larger than the actual value of 
the smaller gap, $\Delta - \delta \mu$. By increasing the value of 
the coupling constant slightly, however, the transitional 2SC phase 
can be made much stronger and the window of intermediate temperatures 
can become considerably wider. In either case, we find it 
rather unusual that the g2SC phase of neutral quark matter is 
replaced by a transitional 2SC phase at intermediate temperatures 
which, at higher temperatures, is replaced by the g2SC phase again. 

\begin{figure}
\epsfxsize=12cm  
\epsffile{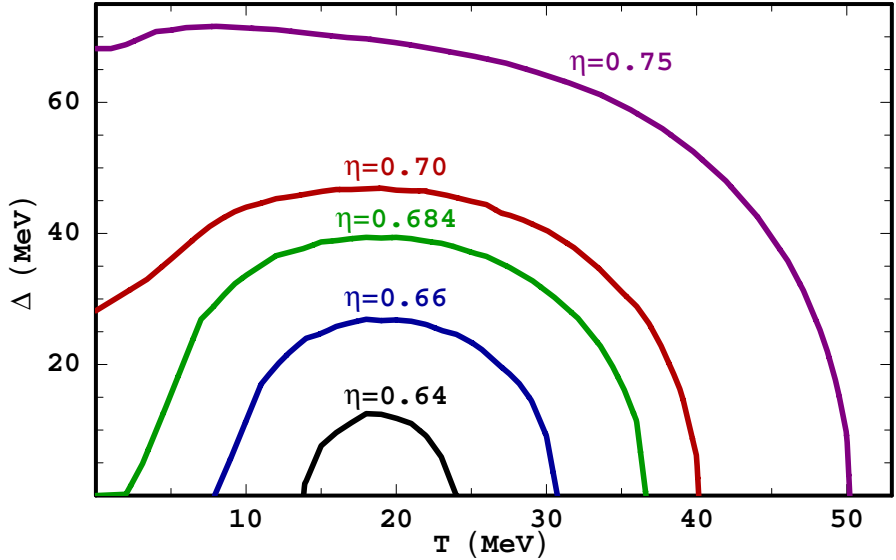}

\caption{The temperature dependence of the diquark gap in neutral 
quark matter calculated for several values of the diquark coupling 
strength $\eta=G_D/G_S$.}

\label{gap-eta}
\end{figure}

It appears that the temperature dependence of the diquark gap is 
very sensitive to the choice of the diquark coupling strength 
$\eta=G_D/G_S$ in the model at hand. This is not surprising because 
the solution to the gap equation is very sensitive to this choice. 
The resulting interplay of the solution for $\Delta$ with the 
condition of charge neutrality, however, is very interesting.
This is demonstrated by the plot of the temperature dependence 
of the diquark gap calculated for several values of the diquark 
coupling in Fig.~\ref{gap-eta}.

%%%%%%%%%%%%%%%%%%%%%%%%%%%%%%%%%%%%%%%%%%%%%%%%%%%%%%%%%%%%%%%%%%%%
%%%%%%%%%%%%%%%%%%%%%%%%%%%%%%%%%%%%%%%%%%%%%%%%%%%%%%%%%%%%%%%%%%%%

The most amazing are the results for weak coupling. It appears that
the gap function could have sizable values at finite temperature even
if it is exactly zero at zero temperature. This possibility comes
about only because of the strong influence of the neutrality condition
on the ground state preference in quark matter. Because of the thermal
effects, the positive electrical charge of the diquark condensate is
easier to accommodate at finite temperature. This opens a possibility 
of the Cooper pairing that is forbidden at zero temperature. 

We should mention that somewhat similar results for the temperature 
dependence of the gap were also obtained in Ref.~\cite{SedLom} in
a study of the asymmetric nuclear matter. The asymmetry parameter 
in nuclear matter $\alpha=(\rho_n-\rho_p)/(\rho_n+\rho_p)$ (where 
$\rho_n$ and $\rho_p$ are the neutron and proton densities, 
respectively) plays a role similar to the coupling strength $\eta$ 
in the quark matter system.

\subsection{Nonuniversal ratio $T_c/\Delta_0$ }

In the preceding subsection, we saw that the temperature dependence of
the gap in the g2SC phase is very different from the standard benchmark 
result in the BCS theory. One of the main properties 
of the BCS temperature dependence is a universal value of the ratio of 
the critical temperature $T_c$ to the value of the gap at zero 
temperature $\Delta_0$, see Eq.~(\ref{r_BCS}). It is instructive, 
therefore, to calculate the same quantity in the g2SC phase of neutral
quark matter. 

Let us start from a simple exercise, and consider the results plotted
in Fig.~\ref{gap-mue0-vs-T.eps} more carefully. First of all, we find 
that $T_c^{2SC} \approx 62.06$ MeV and $\Delta_0^{2SC} = 109.4$ 
MeV when there is no mismatch parameters in the model (i.e., $\mu_e=0$ 
and $\mu_8=0$ represented by dash-dotted line). It is not very 
surprizing that the ratio of interest $r_{2SC}\approx 0.567$ is in 
agreement with the BCS result. Now, if we check the results for the 
gap function in the g2SC phase (solid line), we find that $T_c^{g2SC}
\approx 50.2$ MeV and $\Delta_0^{g2SC} \approx 68.2$ MeV. Thus, the
ratio is $r_{g2SC}\approx0.7357$ which is considerably larger than the 2SC 
result. 

It appears that the real situation is even more interesting. The result 
for the ratio $r_{g2SC}\approx 0.7357$ is not universal. In fact, its 
value depends very much on the diquark coupling constant and, moreover, 
it can even be arbitrarily large. This last statement may not be so 
unexpected if we recall the temperature dependences of the gap shown in 
Fig.~\ref{gap-eta}. 

\begin{figure}
\epsfxsize=12cm  
\epsffile{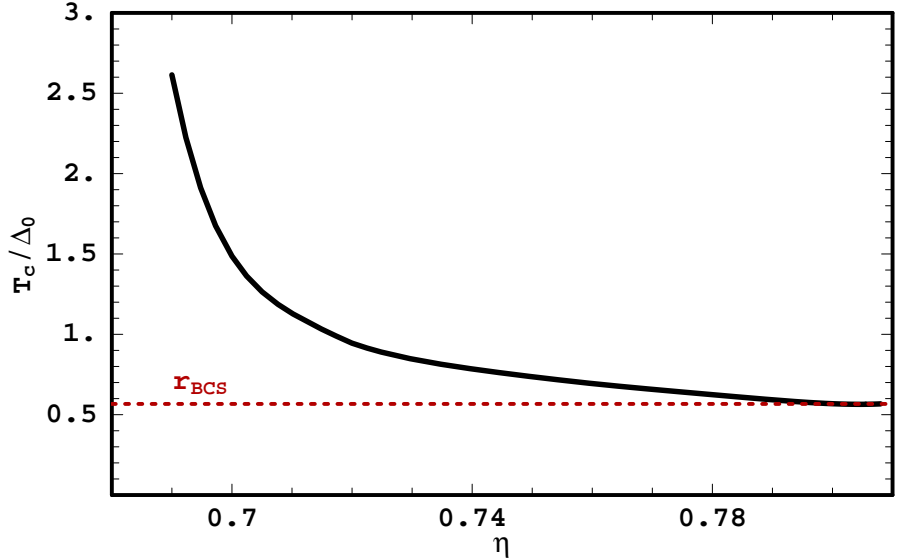}
\caption{The ratio of the critical temperature to the zero temperature 
gap in neutral quark matter as a function of the coupling strength
$\eta=G_D/G_S$.}
\label{ratio}
\end{figure}

The numerical results for the ratio of the critical temperature to 
the zero temperature gap in the g2SC case as a function of the diquark 
coupling strength $\eta=G_D/G_S$ are plotted in Fig.~\ref{ratio}. 
The dependence is shown for the most interesting range of values of 
$\eta=G_D/G_S$, $0.68 \lesssim \eta \lesssim 0.81$, which allows 
the g2SC stable ground state at zero temperature, see the discussion
in Subsec.~\ref{w-i-s}. When the coupling gets weaker in this range,
the zero temperature gap vanishes gradually. As we saw from 
Fig.~\ref{gap-eta}, however, this does not mean that the critical 
temperature vanishes too. Therefore, the ratio of a finite value of 
$T_c$ to the vanishing value of the gap can become arbitrarily 
large. In fact, it remains strictly infinite for a range of 
couplings.

%%%%%%%%%%%%%%%%%%%%%%%%%%%%%%%%%%%%%%%%%%%%%%%%%%%%%%%%%%%%%%%%%%%%
%%%%%%%%%%%%%%%%%%%%%%%%%%%%%%%%%%%%%%%%%%%%%%%%%%%%%%%%%%%%%%%%%%%%

\section{Conclusion}
\label{conclusion}

In this paper, we presented a detailed study of the zero and finite
temperature properties of the gapless color superconducting phase of 
neutral two-flavor quark matter in $\beta$-equilibrium. The spectrum
of quasiparticles in this phase has {\em four} gapless and only two 
gaped fermionic modes, see Eqs.~(\ref{disp-ub})--(\ref{2-degenerate})
with $\Delta<\mu_{e}/2$. This is in contrast to the conventional 2SC 
phase that has only {\em two} gapless and four gaped quasiparticles,
see the case with $\Delta>\mu_{e}/2$. As we show, the choice of the 
ground state is predetermined by the strength of the diquark coupling. 
In strong coupling  regime, the 2SC survives even under the constrain 
of the charge neutrality and $\beta$-equilibrium. When the coupling is 
not very strong, the 2SC ground state is replaced by the g2SC phase. 
Finally, at weak coupling, the stable ground state corresponds to the 
normal phase. 

In the g2SC phase, like in the conventional 2SC phase, the original 
color symmetry SU$(3)_c$ is broken down to SU$(2)_c$. Despite that, 
the low energy spectrum of the gapless phase looks very similar to 
the spectrum in the normal quark matter. By the ``normal quark matter'' 
here we understand the mixture of the following four normal components: 
(i) red and green up quarks with an effective Fermi momentum 
$p_F^{u,{\rm eff}}=\mu^{-}$, (ii) red and green down quarks with 
an effective Fermi momentum $p_F^{d,{\rm eff}}=\mu^{+}$, (iii) 
blue up quarks with $p_F^{ub}=\mu_{ub}$ and (iv) blue down quarks 
with $p_F^{db}=\mu_{db}$. Needless to say that the similarity
extends only in a small region of low energies, $E\lesssim 
\delta\mu-\Delta$. This similarity suggests that there is no 
color Meissner effect in this phase of matter, although the gauge 
symmetry is broken through the Anderson-Higgs mechanism. 

The same arguments regarding the resemblance of the low energy 
spectrum in the g2SC and normal phases of quark matter lead us to 
the conclusion that two ``secondary'' spin-1 condensates should 
develop on top of the gapless ground state. Indeed, the existence 
of highly degenerate gapless modes around the effective two Fermi 
surfaces should lead to the appearance of additional condensates, 
provided an attractive channel exists. Because of the red-green 
color degeneracy of the quasiparticles around $p_F^{u,{\rm eff}}
=\mu^{-}$ and $p_F^{d,{\rm eff}}=\mu^{+}$, the attractive 
antisymmetric channel should exist. As a result, two spin-1 gaps 
would open around the points where the gapless modes used to be.

Most likely, the appearance of the spin-1 gaps is a small effect on 
top of the g2SC ground state. Therefore, many thermodynamical properties 
of quark matter would not be affected by their presence. The transport 
properties, on the other hand, should be modified. In particular, five 
gluons of the original SU$(3)_c$ gauge group should experience the 
color Meissner effect. In addition there will also be the electromagnetic 
Meissner effect \cite{spin-1-Meissner}.

At finite temperature, the properties of the g2SC phase are very 
unusual. One of the most striking results is that the ratio of 
the critical temperature to the gap at zero temperature is not a 
universal number. It depends on the choice of the coupling constant
in the diquark channel. Moreover, the value of the ratio can become 
arbitrarily large with proper choice of the coupling. For a range
of values, it is infinite. This is a simple consequence of the 
nonmonotonic temperature dependence of the diquark gap function 
on temperature. In an extreme case, the finite temperature gap 
can be nonzero even if the zero temperature gap is zero. To the 
best of our knowledge, the only other example of such a temperature 
behavior of an order parameter was obtained in Ref.~\cite{SedLom} 
in a study of the asymmetric nuclear matter.

In nature, gapless 2SC quark matter could exist in compact stars. 
Indeed, this phase is neutral with respect to electric and color 
charges and satisfies the $\beta$-equilibrium condition by construction. 
If this is indeed the case, one may have a chance to detect the 
indirect signatures of its presence by deciphering the observational 
data from stars. The study of the physical properties of the g2SC
phase in this paper is the first step in this direction. In future,
one should also study the transport properties of the g2SC phase.
For example, the knowledge of the neutrino emissivities and mean 
free paths would be crucial for the understanding of the cooling 
rates of stars with such a gapless phase in their interior. 

As we mentioned earlier, some stable gapless phases may also be 
realized in the asymmetric nuclear matter \cite{SedLom} and in 
two-component mixtures of cold fermionic atoms \cite{WilLiu}. 
The results of this paper might also be useful for a deeper
understanding of the properties of such nonrelativistic systems. 
For example, the similarity of the spectrum of the low energy 
quasiparticles in the gapless phase and the normal matter seems
to suggest that there is no superfluidity in the corresponding 
phases of cold atom mixtures. Of course, one should be careful 
before making a conclusive statement because the corresponding 
nonrelativistic systems are quite different from neutral quark 
matter.

{\bf Acknowledgements}. The authors thank Prof. D.H.~Rischke for
stimulating discussions as well as H.~Caldas and P.~Bedaque for 
comments. I.A.S. would like to thank T.~Sch{\"a}fer for discussions. 
M.H. acknowledges the financial support from 
the Alexander von Humboldt-Foundation, and the NSFC under Grant Nos. 
10105005, 10135030. The work of I.A.S. was supported by Gesellschaft 
f\"{u}r Schwerionenforschung (GSI) and by Bundesministerium f\"{u}r 
Bildung und Forschung (BMBF).

\end{document}